\documentclass[a4paper]{article}

\usepackage{aas_macros}
\usepackage{graphicx}%
\usepackage{multirow}%
\usepackage{amsmath,amssymb,amsfonts}%
\usepackage{amsthm}%
\usepackage{mathrsfs}%
\usepackage[title]{appendix}%
\usepackage{xcolor}%
\usepackage{textcomp}%
\usepackage{manyfoot}%
\usepackage{booktabs}%
\usepackage{algorithm}%
\usepackage{algorithmicx}%
\usepackage{algpseudocode}%
\usepackage{listings}%
\usepackage{ulem}
\usepackage{lscape}
\usepackage{units}
\usepackage{rotating}
\usepackage{hyperref}
\usepackage{tablefootnote}
\usepackage{amssymb,amsmath,caption}

\usepackage[
    backend=biber,
    bibencoding=utf8,
    style=numeric, 
    citestyle=numeric-comp,
    firstinits=true,
    isbn=false,
    doi=true,
    url=false,
    defernumbers=true,
    clearlang=true,
    sorting=none,
    sortcites,
]{biblatex}

\defbibfilter{onlymain}{%
  segment=0
}

\defbibfilter{onlymethods}{%
  not segment=0
}

\usepackage{authblk}

\author[1,2]{Roberta Tripodi \thanks{Corresponding Author: roberta.tripodi@fmf.uni-lj.si}}
\author[1]{Nicholas Martis}
\author[1]{Vladan Markov}
\author[1]{Maru\v{s}a Brada\v{c}}
\author[3]{Fabio Di Mascia}
\author[4]{Vieri Cammelli}
\author[5,6]{Francesco D'Eugenio}
\author[7]{Chris Willott}
\author[8]{Mirko Curti}
\author[3]{Maulik Bhatt}
\author[3]{Simona Gallerani}
\author[1]{Gregor Rihtar\v{s}i\v{c}}
\author[9]{Jasbir Singh}
\author[10]{Gaia Gaspar}
\author[1]{Anishya Harshan}
\author[1]{Jon Jude\v{z}}
\author[10]{Rosa M. Merida}
\author[10,11]{Guillaume Desprez}
\author[10]{Marcin Sawicki}
\author[12]{Ilias Goovaerts}
\author[13]{Adam Muzzin}
\author[12]{Ga\"el Noirot}
\author[13]{Ghassan T.E. Sarrouh}
\author[14,15]{Roberto Abraham}
\author[10,16]{Yoshihisa Asada}
\author[17]{Gabriel Brammer}
\author[10]{Vicente Estrada-Carpenter}
\author[1]{Giordano Felicioni}
\author[14,18]{Seiji Fujimoto}
\author[19]{Kartheik Iyer}
\author[20]{Lamiya Mowla}
\author[21,22]{Victoria Strait}

\affil[1]{Faculty of Mathematics and Physics, University of Ljubljana, 19 Jadranska ulica, Ljubljana, 1000, Slovenia}

\affil[2]{IFPU, Institute for Fundamental Physics of the Universe, Via Beirut 2, Trieste, 34151, Italy}

\affil[3]{Scuola Normale Superiore, Piazza dei Cavalieri, 7 Pisa, 56126, Italy}

\affil[4]{Astronomy Unit, Department of Physics, University of Trieste, via G.B. Tiepolo 11, Trieste, 34143, Italy}

\affil[5]{Kavli Institute for Cosmology, University of Cambridge, Madingley Road, CB3 0HA, Cambridge, UK}

\affil[6]{Cavendish Laboratory - Astrophysics group, University of Cambridge, 19 JJ Thomson Avenue, Cambridge, CB30HE, UK}

\affil[7]{NRC Herzberg, 5071 West Saanich Rd, Victoria, BC V9E 2E7, Canada}

\affil[8]{European Southern Observatory, Karl-Schwarzschild-Strasse 2, Garching, 85748, Germany}

\affil[9]{INAF - Osservatorio Astronomico di Brera, via Brera 20, I-20121, Milano, Italy}

\affil[10]{Department of Astronomy and Physics and Institute for Computational Astrophysics, Saint Mary's University, 923 Robie Street, Halifax, B3H 3C3, Nova Scotia, Canada}

\affil[11]{Kapteyn Astronomical Institute, University of Groningen, P.O. Box 800, Groningen, 9700AV, The Netherlands}

\affil[12]{Space Telescope Science Institute, 3700 San Martin Drive, Baltimore, MD 21218, USA}

\affil[13]{Department of Physics and Astronomy, York University, 4700 Keele St., Toronto, M3J 1P3, Ontario, Canada}

\affil[14]{David A. Dunlap Department of Astronomy and Astrophysics, University of Toronto, 50 St. George Street, Toronto, Ontario, M5S 3H4, Canada}

\affil[15]{Dunlap Institute for Astronomy and Astrophysics, 50 St. George Street, Toronto, Ontario, M5S 3H4, Canada}

\affil[16]{Department of Astronomy, Kyoto University, Sakyo-ku, Kyoto, 606-8502, Japan}

\affil[17]{Niels Bohr Institute, University of Copenhagen, Jagtvej 128, DK-2200, Copenhagen, Denmark}

\affil[18]{Department of Astronomy, The University of Texas at Austin, Austin, TX 78712, USA}

\affil[19]{Columbia Astrophysics Laboratory, Columbia University, 550 West 120th Street, New York, 10027, NY, USA}

\affil[20]{Whitin Observatory, Department of Physics and Astronomy, Wellesley College, 106 Central Street, Wellesley, MA 02481, USA}

\affil[21]{Cosmic Dawn Center (DAWN),Denmark}

\affil[22]{Niels Bohr Institute, University of Copenhagen, Jagtvej 128, Copenhagen, DK-2200, Denmark}

\date{}

\makeatletter
\let\blx@rerun@biber\relax
\makeatother

\def\be{\begin{eqnarray}}
\def\ee{\end{eqnarray}}

\usepackage{textgreek}
\usepackage{xargs}
\usepackage{xcolor}
\usepackage{xspace}

\let\oldtextsigma\textsigma
\renewcommand{\textsigma}{\oldtextsigma\xspace}
\let\oldAA\AA
\renewcommand{\AA}{\text{\oldAA}\xspace}
\def\w80{\ensuremath{w_{80}}\xspace}

\newcommandx{\fluxdcgs}[1][1=-20]{$\times 10^{[#1]}$~erg~s$^{-1}$~cm$^{-2}$~\AA$^{-1}$\xspace}

\newcommand{\Hbeta}{\text{H\textbeta}\xspace}
\newcommand{\Hgamma}{\text{H\textgamma}\xspace}
\newcommand{\Hdelta}{\text{H\textdelta}\xspace}
\newcommand{\Hzeta}{\text{H\textzeta}\xspace}
\newcommand{\Heta}{\text{H\texteta}\xspace}
\newcommandx{\permittedEL}[6][1=O,2=III,3=,4=,5=,6=]{\text{{#1}\,{\sc{#2}}{#3}{#4}{#5}{#6}}\xspace}
\newcommandx{\semiforbiddenEL}[6][1=O,2=III,3=,4=,5=,6=]{\text{{#1}\,{\sc{#2}}]{#3}{#4}{#5}{#6}}\xspace}
\newcommandx{\forbiddenEL}[6][1=O,2=III,3=,4=,5=,6=]{\text{[{#1}\,{\sc{#2}}]{#3}{#4}{#5}{#6}}\xspace}

\newcommandx{\HII}{\permittedEL[H][ii]}
\newcommandx{\HeI}{\permittedEL[He][i]}
\newcommandx{\HeIL}[1][1=3889]{\permittedEL[He][i][\,\textlambda][#1]}
\newcommandx{\HeIIL}[1][1=4686]{\permittedEL[He][ii][\,\textlambda][#1]}
\newcommand{\OIII}{\forbiddenEL[O][iii]}
\newcommandx{\OIIIL}[1][1=5007]{\forbiddenEL[O][iii][\textlambda][#1]}
\newcommand{\OIIIall}{\forbiddenEL[O][iii][\textlambda][\textlambda][4959,][5007]}
\newcommandx{\OIIIlow}[1][1=1666]{\semiforbiddenEL[O][iii][\textlambda][#1]}
\newcommandx{\NIL}[1]{\forbiddenEL[N][i][\textlambda][5200]}

\newcommand{\OII}{\forbiddenEL[O][ii]}
\newcommandx{\OIIL}[1][1=3727]{\forbiddenEL[O][ii][\textlambda][#1]}

\newcommand{\OIIalllow}{\forbiddenEL[O][ii][\textlambda][\textlambda][][7322,7332]}

\newcommand{\NeIII}{\forbiddenEL[Ne][iii][\textlambda][3869]}
\newcommand{\NeIIIe}{\forbiddenEL[Ne][iii]}

\newcommandx{\NIIL}[1][1=6583]{\forbiddenEL[N][ii][\textlambda][#1]}
\newcommand{\NeIIIb}{\forbiddenEL[Ne][iii][\textlambda][3967]}

\newcommand{\NIV}{\semiforbiddenEL[N][iv][\textlambda][\textlambda][1483,][1486]}
\newcommandx{\CIVall}{\permittedEL[C][iv][\textlambda][\textlambda][1549,][1551]}
\newcommandx{\CIV}{\permittedEL[C][iv]}

\newcommand{\CIII}{\semiforbiddenEL[C][iii]}
\newcommandx{\NV}{\permittedEL[N][v]}
\newcommandx{\NeV}{\forbiddenEL[Ne][v]}

\begin{document}
\title{\huge Red, hot, and very metal poor: extreme properties of a massive accreting black hole in the first 500 Myr}

\maketitle



\begin{refsegment}

\textbf{The James Webb Space Telescope (JWST) has recently discovered a new population of objects at high redshift referred to as `Little Red Dots' (LRDs). Their nature currently remains elusive \cite{gonzalez2024}, despite their surprisingly high inferred number densities \cite{williams2024,matthee2024,greene2024, kocevski2024,kokorev2024,hainline2024, maiolino2023}. This emerging population of red point-like sources is reshaping our view of the early Universe and may shed light on the formation of high-redshift supermassive black holes. Here we present a spectroscopically confirmed LRD CANUCS-LRD-z8.6 at $z_{\rm spec}=8.6319\pm 0.0005$ hosting an Active Galactic Nucleus (AGN), using JWST data. This source shows the typical spectral shape of an LRD (blue UV and red optical continuum, unresolved in JWST imaging),  along with broad \Hbeta line emission, detection of high-ionization emission lines (\CIV, \NIV) and very high electron temperature indicative of the presence of AGN. This is also combined with a very low metallicity ($Z<0.1 Z_\odot$). The presence of all these diverse features in one source makes CANUCS-LRD-z8.6 unique. We show that the inferred black hole mass of CANUCS-LRD-z8.6 ($M_{\rm BH}=1.0^{+0.6}_{-0.4}\times 10^{8}\rm ~M_\odot$) strongly challenges current standard theoretical models and simulations of black hole formation, and forces us to adopt `ad hoc' prescriptions. Indeed if massive seeds, or light seeds with super-Eddington accretion, are considered, the observed BH mass of CANUCS-LRD-z8.6 at $z=8.6$ can be reproduced. Moreover, the black hole is over-massive compared to its host, relative to the local $M_{\rm BH}-M_*$ relations, pointing towards an earlier and faster evolution of the black hole compared to its host galaxy.} 


%
%
%





\bigskip

\begin{figure}[h]
    \centering
    \includegraphics[width=0.9\linewidth]{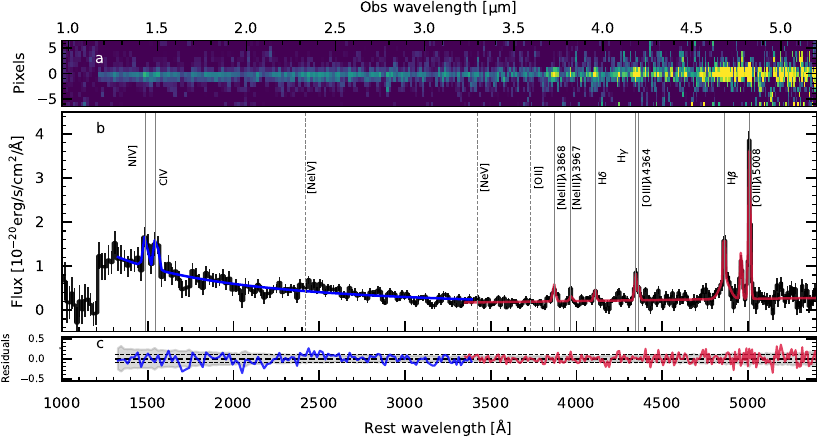} \\
    \includegraphics[width=0.9\linewidth]{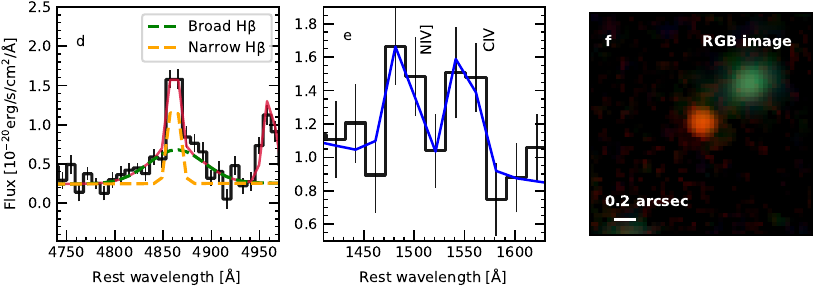}
    \caption[]{{\bf NIRSpec PRISM spectrum and RGB image of CANUCS-LRD-z8.6 at $\mathbf{z= 8.6319\pm 0.0005}$}. Panel a: 2D spectrum. Panel b: The black line represents the measured 1D spectrum. The two independent best-fitting functions for the rest-frame UV and optical part are shown as blue and red solid curves, respectively. Vertical solid (dashed) lines mark detected (non-detected) emission lines, respectively. Panel c: residuals of the fit of the 1D spectrum. The dashed horizontal lines mark the 0 level and the average $\pm 1\sigma$ noise levels derived at $\lambda>2000\AA$. The grey shaded area is $1\sigma$ rms. Panel d: Zoom in on the \Hbeta emission line. The best-fitting spectrum is shown in red, while the narrow and broad \Hbeta components are shown in yellow and green, respectively. Panel e: zoom-in of the \NIV and \CIV emission lines. The best-fitting spectrum is shown in blue. Panel f: RGB image of CANUCS-LRD-z8.6, combining psf-matched imaging using F090W, F200W, and F444W filters. The scale of the image is reported in the lower left corner.}
    \label{fig:spectra}
\end{figure}

LRDs appear as a heterogeneous galaxy population in which both active galactic nuclei (AGN) and star formation can contribute to their observed light. Some LRDs have been classified as AGN hosts based on the detection of broad emission in Balmer lines \cite{kokorev2023,matthee2024,taylor2024}, and some show evidence for evolved stellar populations with clear Balmer breaks \cite{wang2024}. Moreover, it has been recently found that, independent of the specific AGN contribution adopted, the LRDs’ black holes (BHs) are significantly overmassive relative to their host galaxies compared to the local $M_{\rm BH}-M_*$ relation \cite{durodola2024}, and their formation channels remain unclear. Understanding how these massive BHs formed in such compact galaxies as early as redshift $z = 8.6$ remains a key question \cite{kokorev2023,silk2024}. In this context, the discovery of the LRD CANUCS-LRD-z8.6 at $z_{\rm spec}=8.6319\pm 0.0005$ (object ID: 5112687) opens a new pathway to the understanding of this intriguing population of LRDs. Indeed, this source is unique in terms of its BH and host galaxy properties, being the only high-z source to date to show evidence of broad line emission and high-ionization lines.

As part of the CAnadian NIRISS Unbiased Cluster Survey (CANUCS) program, JWST/NIRCam and NIRSpec observations of the cluster MACS J1149.5+2223 allowed us to identify and study CANUCS-LRD-z8.6 (Figure \ref{fig:spectra}). The galaxy was first selected as a high-redshift `double break' galaxy \cite{desprez2024}, and it also meets the criteria set by ref. \cite{kocevski2024} on spectral slopes and size to be classified as an LRD. The upper limit derived for the half-light radius of this source is very stringent, $r<70$ pc (see Methods).

As shown in Figure \ref{fig:spectra}.d, the detection of broad emission in \Hbeta (S/N$=6$), with ${\rm FWHM}_{\rm \Hbeta, broad}=4200^{+600}_{-500} ~\rm km ~s^{-1}$ (see Methods), indicates the presence of an AGN. 
The non-detection of broad emission in \OIIIL[5008] refutes the scenario of the high gas velocities arising from outflowing material. From locally-calibrated empirical relations \cite{greene2005} using both the broad \Hbeta luminosity and the $L_{5100\AA}$ continuum luminosity, we derive consistent values for the BH mass of $M_{\rm BH}=1.0^{+0.6}_{-0.4} \times 10^8 \rm ~M_\odot$. This picture is further supported by the detection of high ionization lines of \NIV  and \CIV (both at S/N\,$=3$; Figure \ref{fig:spectra}e). These are typical indicators of Type II AGNs \cite{feltre2016}, and have been seen with JWST in galaxies identified as Type II AGNs \cite{scholtz2023,treiber2024}. For CANUCS-LRD-z8.6, the detection of both \NIV  and \CIV cannot be explained as arising from star formation \cite[as in][, see Methods]{topping2024}. Moreover, the combination of extremely high-velocity broad line gas ($>1000$ km s$^{-1}$) and high-ionization narrow emission lines with high photo-ionization energy ($\sim 50-60$ eV) constitutes a strong AGN signature \cite{mazzucchelli2023, dodorico2023, kokorev2023}. When compared with other spectroscopically well-studied objects at $z>7$ \cite{bunker2023, maiolino2024,deugenio2024,castellano2024,kokorev2023,topping2024,ubler2024,stark2015}, 
CANUCS-LRD-z8.6 stands out given that it is the only source showing both very broad emission in \Hbeta and the detection of \NIV and \CIV. Notably, compared to the AGN candidate GN-z11 at $z = 10.6$, which also exhibits \NIV and \CIV emission \cite{bunker2023, maiolino2024}, CANUCS-LRD-z8.6 hosts a supermassive black hole (SMBH) approximately 100 times more massive than that of GN-z11, despite being just 150 Myr older.    

\begin{figure}[t]
    \centering
    \includegraphics[width=1\linewidth]{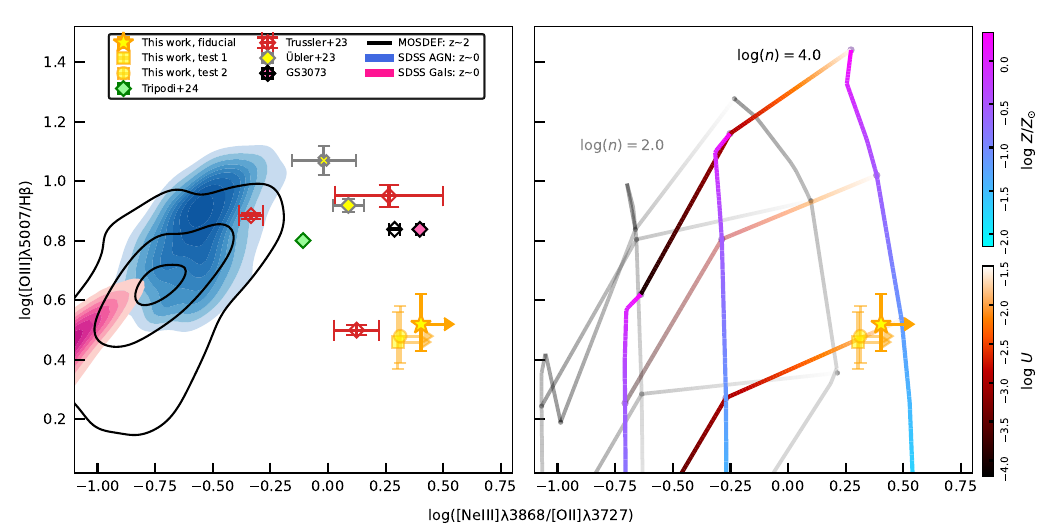}
    \caption[]{{\bf \OIIIL[5007]/\Hbeta--\NeIII/\OIIL[3727] narrow line ratio diagram}. The fiducial result for CANUCS-LRD-z8.6 is shown as a yellow star. The lighter yellow square and circle are the results for the line ratios of CANUCS-LRD-z8.6 from Test 1 and Test 2, respectively, derived by adding a broad \Hgamma component (see Methods). Line ratios are dust-corrected. The maximum correction for \NeIII/(\NeIII+\Hzeta+\Heta) due to the blend of \NeIII, \HeIL, \Heta and \Hzeta is $\approx$0.2 dex (see Methods). Left: the \OIIIL[5007]/\Hbeta ratio as a function of the \NeIII/\OIIL[3727] ratio (`OHNO' diagnostic) for our target and other observations as described in the legend and in the Methods. Right: overlaid to our results for the `OHNO' diagnostic are the AGN photoionization models of \cite{feltre2016} at hydrogen densities $\log n{\rm [cm^{-3}]}=2.0$ (grey scale) and $\log n{\rm [cm^{-3}]}=4.0$ (coloured scale). The grid shows the variation of the ionization parameter and metallicity (colour scales on the right-hand side of the figure, same limits are adopted for gray scale). To guide the eye, lighter colors represent lower metallicity and higher ionization.}
    \label{fig:diagnostic}
\end{figure}

As anticipated above, the properties of CANUCS-LRD-z8.6's host galaxy also add interesting details to the picture, specifically in terms of the composition of the gas and the co-evolution between the SMBH and its host. Since the auroral \OIIIL[4364] emission line has been detected with S/N$=4$, albeit blended with a stronger \Hgamma line, we can determine the electron temperature ($T_e$) and the gas-phase metallicity. We find an extreme \OIIIL[4364]/\OIIIL[5008] ratio ($\sim -1$ dex) even after subtracting an additional broad \Hgamma component (see Methods). These ratios are found to be suggestive of high densities as in the densest part of the narrow line region of Type I AGNs \cite{binette2024}. Using \texttt{pyneb} models, we derive $T_e=40,000_{-12,000}^{+16,000}$ K, which is 2-4 times higher than in star-forming galaxies \cite{curti2023,curti2024b}, but consistent with other AGNs \cite{baskin2005,kokorev2023}. Even when considering the maximum \Hgamma contribution, the electron temperature is at least $>20,000$ K. Such a high electron temperature further supports the presence of an AGN in CANUCS-LRD-z8.6. Moreover, when comparing  the \OIIIL[4364]/\Hgamma versus \OIIIL[5008]/\OIIIL[4364] line ratio diagram as done in \cite{mazzolari2024,ubler2024}, our target occupies the region of higher electron temperatures and displays elevated ratios of \OIIIL[4364]/\Hgamma compared to local AGNs, similar to the $z=7.15$ AGN type I's host galaxy ZS7, identified as a broad-line AGN \cite{ubler2024}. The stringent upper limit on the \OII[3727] emission line provides strong constraints on the metallicity content of this LRD. Indeed, the $3\sigma$ upper limit inferred from the $T_e$-direct method is low, $Z\lesssim 0.2 ~Z_\odot$ (see Methods). 

\begin{figure}[t]
    \centering
    \includegraphics[width=0.8\linewidth]{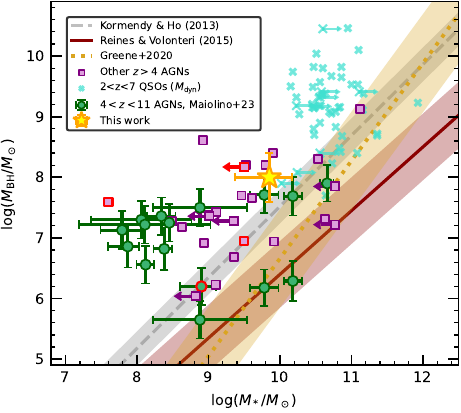}
    \caption[]{{\bf Black hole mass versus stellar mass.} The results for CANUCS-LRD-z8.6 (yellow star) are compared with $4<z<11$ AGNs (green dots) from ref. \cite{maiolino2023}, a compilation of AGNs at $z>4$ (purple squares) discovered by other JWST surveys at high-z \cite{harikane2023,larson2023,ubler2023,kocevski2023,ding2023,bogdan2024,kokorev2023,juodzbalis2024,wang2024,akins2024}, and with local scaling relations \cite[dashed grey, solid red and dotted yellow lines, respectively from refs.][]{kormendy2013,reines2015,greene2020}. Red edges mark sources at $z>8$ \cite[][]{kokorev2023,larson2023,bogdan2024,maiolino2024}.}
    \label{fig:mbh-mstar}
\end{figure}

These results are also consistent with those derived from `OHNO' diagnostic, shown in Figure \ref{fig:diagnostic}, which relates the \OIII[5008]/\Hbeta with \NeIII/\OII[3727] to distinguish between the star-forming or AGN nature of galaxies at low- and high-z \cite{backhaus2022, cleri2022, larson2023}. Also in this case, the observed `OHNO' ratios (yellow star) clearly indicate low metal content ($Z\lesssim 0.1 ~Z_\odot$) and high ionization parameter ($\log(U)\sim -1.5$) when comparing with photo-ionization models.

 By performing Spectral Energy distribution (SED) fitting using both photometry and spectroscopy with \texttt{Bagpipes} \cite{carnall2018, carnall2019}, including an AGN component, we derive a stellar mass of $M_*= 7.6^{+7.2}_{-5.2}\times 10^9 \rm ~M_\odot$ (errors include possible variation of the SED and dust modelling, see Methods).  As shown in Figure \ref{fig:mbh-mstar}, in terms of stellar mass, CANUCS-LRD-z8.6 is indeed the most extreme AGN at $z>8$.  
 
 LRDs' BHs are found to be significantly overmassive relative to their host galaxies compared to the local $M_{\rm BH}-M_{*}$ relation \cite{durodola2024}. Given the SED-inferred $M_*$, it is possible to place CANUCS-LRD-z8.6 on the $M_{\rm BH}-M_{*}$ plane. As presented in Figure \ref{fig:mbh-mstar}, CANUCS-LRD-z8.6 stands above the $M_{\rm BH}-M_{*}$ local relations (in particular the one for local AGNs, in red; \cite{reines2015}), indicating faster evolution of the BH compared to its host galaxy. We are witnessing the growth of a SMBH of $10^{8}~\rm M_\odot$ in a very compact and massive galaxy ($M_*\simeq 8 \times 10^9~\rm M_\odot$ in $r<70$ pc), unlike any other sources at the same redshift. The SMBH in CANUCS-LRD-z8.6 has a comparable mass to that of GS-9209 at $z \sim 4.7$, with the critical difference that GS-9209 is quenched, having halted star formation by $z \sim 7$ \cite{carnall2023}. Additionally, the stellar mass of CANUCS-LRD-z8.6 appears to be approaching that of the massive post-starburst galaxy RUBIES-UDS-QG-z7 ($M_*\sim 1.7\times 10^{10} ~\rm M_\odot$, \cite{weibel2024b}), which is already quiescent at $z \sim 7$. These observations suggest that SMBHs as massive as CANUCS-LRD-z8.6 could play a pivotal role in quenching star formation in galaxies by $z \sim 7$.
 
 \begin{figure}[t]
 	\centering
 	\includegraphics[width=0.8\linewidth]{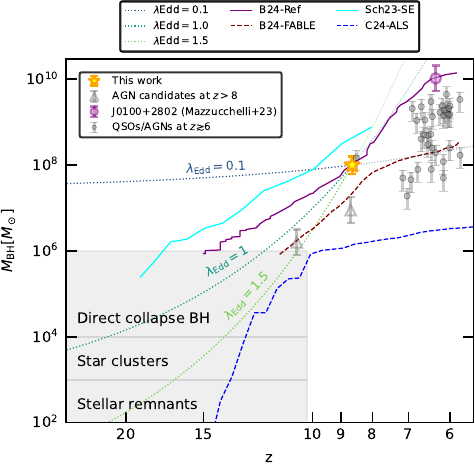}
 	\caption[]{{\bf Black hole mass accretion history.} The growth history of CANUCS-LRD-z8.6 assuming a constant accretion rate corresponding to the observed luminosity ($\lambda_{\rm Edd}=0.1$), Eddington capped accretion ($\lambda_{\rm Edd}=1$), and super-Eddington accretion ($\lambda_{\rm Edd}=1.5$), as blue, teal and light green dotted lines, respectively. Solid (dashed) lines are simulations \cite{bennett2024} or semi-analytical models \cite{Cammelli24,Schneider2023} with different prescriptions, as detailed in the Methods, that can (cannot) reproduce the BH mass of CANUCS-LRD-z8.6. We show for comparison only QSOs and AGNs having BH masses derived from broad lines \cite[as gray dots,][]{mazzucchelli2023,maiolino2023,kokorev2023,tripodi2024,larson2023,juodzbalis2024,ubler2024,akins2024}, and two AGN candidates at $z>6$ \cite[as gray triangles,][]{larson2023,maiolino2024}.}
 	\label{fig:bh_growth}
 \end{figure}
 
The high BH mass of CANUCS-LRD-z8.6 at $z=8.6$ imposes stringent constraints on the formation pathways of SMBHs, challenging standard models and simulations. When examining potential growth trajectories under constant accretion rates (dotted lines in Figure \ref{fig:bh_growth}), the mass of CANUCS-LRD-z8.6's BH suggests an origin in massive seed mechanisms \cite{bogdan2024, huang2024}, possibly from direct collapse or Pop III.1\footnote{Pop III.1 stars are a subclass of Population III stars, which ref. \cite{MT08} divided into two categories. Pop III.1 stars are a unique type of Population III stars that form at the centers of dark matter minihalos in the early universe ($z\gtrsim20$), remaining isolated from any stellar or BH feedback \cite{MT08}. In contrast, Pop III.2 stars also form within dark matter minihalos but are affected by feedback from external astrophysical sources. This external influence promotes gas fragmentation, leading to the formation of lower-mass stars compared to Pop III.1.} models \cite[e.g.,][]{Bromm03, Maio19, Bhowmick22a, Singh23, Cammelli24}. While lower-mass seeds are easier to account for, sustaining a constant or super-Eddington accretion rate over the BH's history has been largely ruled out by numerous studies \cite[e.g.,][]{Oshea05, Jeon23}. We refer to the Methods for a more exhaustive comparison among semi-analytical models (SAMs) and numerical simulations with different prescriptions, and we summarize here our main findings. In SAMs, Super-Eddington accretion results to be essential to assemble a sufficiently large amount of mass within 500 Myr (cyan solid line in Figure \ref{fig:bh_growth}). In fact, SAMs assuming Eddington limited growth, though matching luminosity functions up to $z \sim 9$ and the $M_{\text{BH}}-M_{*}$ relation at $z=0$ (blue lines, \cite{Cammelli24}), fail to account for CANUCS-LRD-z8.6 by two orders of magnitude. Most numerical simulations employing standard prescriptions cannot replicate CANUCS-LRD-z8.6’s BH mass, since AGN feedback prevents early BH growth. In particular, the original recipe of FABLE\footnote{The Feedback Acting on Baryons in Large-scale Environments (FABLE) suite of cosmological hydrodynamical simulations are based upon the framework of the successful Illustris project but improve upon the agreement with observations on scales larger than galaxies \cite{henden2018}.} zoom-in simulations falls short by one order of magnitude (purple dotted track in Figure \ref{fig:bh_growth}). However, modifications promoting earlier BH growth to form BHs of $M_{\rm BH}=10^{10}~\rm M_\odot$ at $z \sim 6$ \cite{bennett2024}, such as QSO J0100+2802 \cite{mazzucchelli2023}, allow to explain the existence of CANUCS-LRD-z8.6 (purple solid track in Figure \ref{fig:bh_growth}; see Methods). As such, CANUCS-LRD-z8.6 is a possible progenitor for QSOs like J0100+2802, which also challenges standard theoretical models. If there exists a substantial population of such SMBHs at $z\sim 8-9$ comprising CANUCS-LRD-z8.6, the highest-redshift example hitherto detected, along with UNCOVER 20466 \cite{kokorev2023}, our understanding of early galaxy evolution and its link to the local Universe may need substantial revision. 

 

\printbibliography[segment=1]
\end{refsegment}

\bigskip
\subsection*{Acknowledgments}
 RT, MB, NM, VM, AH, GR, JJ, and GF acknowledges support from the ERC Grant FIRSTLIGHT and from the Slovenian national research agency ARRS through grants N1-0238, P1-0188 and the program HST-GO-16667, provided through a grant from the STScI under NASA contract NAS5-26555. This research was enabled by grant 18JWST-GTO1 from the Canadian Space Agency, and Discovery Grants from the Natural Sciences and Engineering Research Council of Canada to MS, RA, and AM. YA is supported by a Research Fellowship for Young Scientists from the Japan Society of the Promotion of Science (JSPS) and by the JSPS International Leading Research (ILR) project (KAKENHI Grant Number JP22K21349).
 
\subsection*{Authors contribution} RT led the writing of this paper, performed the analysis of the spectrum, and derived the majority of the physical quantities discussed in this work. RT, YA, CW, GD, contributed to the target selection. CW, NM, VS, GB contributed to reduce the images. GS, YA, AM, CW contributed to the photometry measurements. CW, GB contributed to reduce the NIRSpec data. GS contributed to the psf-matching. NM contributed to the \texttt{galfit} modelling. RT, NM, VM, MBr, FDE, CW, RM, GG, AH contributed to the analysis and discussion of the spectro-photometric spectral energy distribution modelling. RT, FDE, MC contributed to the discussion of the gas properties of the target. FDE contributed to the comparison with photoionization models. MC contributed to the estimate of the metallicity of the target. RT, NM, VM contributed to the discussion of the obscuration of the target. FDM, SG, VC, MB, JS, contributed to the comparison with simulations and semi-analytical models. GR contributed to the lensing modelling and the estimate of the target magnification. JJ contributed to the size estimate of the target. NM, VM, MBr, VC, FDE, CW, FDM, GR, JS, JJ contributed to edit the paper. AH, GD, MS, IG, AM, GN, GS, RA, YA, GF, SF, KI, VS provided comments on the physical interpretation of the results. All authors reviewed the manuscript.

\subsection*{Competing interests} The authors declare no competing interests.

\subsection*{Correspondence and requests for materials} should be addressed to R.T.

\clearpage


\clearpage

\begin{refsegment}

\section{METHODS}\label{sec:meth}

\subsection{Observations and data reduction.}
\label{sec:data}

We use the data acquired by the CANUCS NIRISS GTO Program \#1208 \cite{willott2022} which covers five different strong
lensing cluster (CLU) fields: Abell 370, MACS J0416.1-2403, MACS J0417.5-1154, MACS J1149.5+2223 (hereafter MACS1149, $z=0.543$), and MACS J1423.8+2404 \cite{soucail1987, ebeling2001}. As NIRCam and NIRISS are operated in parallel, each of the five cluster fields is accompanied by a NIRCam flanking field and a NIRISS flanking field. 
Our target lies in the MACS1149 cluster field. 
MACS1149 was observed with NIRCam filters: F090W, F115W, F150W, F200W, F277W, F356W, F410M, and F444W with exposure times of 6.4 ks each. 
We also utilized archival data of {\tt HST} imaging from HFF program \cite{lotz2017}. The CANUCS image reduction and photometry procedure is described in detail in refs. \cite{noirot2023, willott2024, asada2024}, while details on point spread function (PSF) measurement and homogenization can be found in \cite{sarrouh2024}. Cluster galaxies and intra-cluster light were modelled and subtracted to avoid contamination of the photometry as described in ref. \cite{martis2024}. In short, we use a modified version of the Detector1Pipeline ({\tt calwebb\textunderscore detector1}) stage of the official STScI pipeline and jwst\textunderscore 0916.pmap JWST Operational Pipeline (CRDS\textunderscore CTX) to reduce the NIRCam data. We perform astrometric alignment of the different exposures of JWST/NIRCam to HST/ACS images, sky subtraction, and drizzling to a common pixel scale of 0.04$''$ using version 1.6.0 of the grism redshift and line analysis software for space-based slit-less spectroscopy (\texttt{Grizli} \cite{brammer2021}). PSFs are extracted empirically by median stacking bright, isolated, non-saturated stars following the methodology described in ref. \cite{sarrouh2024}. All images are then degraded to the F444W resolution for photometry. The source detection and photometry are done with the Photutils package \cite{bradley2022} on the $\chi_{mean}$ detection image created using all available NIRCam images. 

First selected as a high-z double break galaxy \cite{desprez2024}, our target is classified as an LRD following the criteria on UV and optical slopes and compactness given by ref. \cite{kocevski2024}.

The CANUCS program also includes NIRSpec low-resolution prism multi-object spectroscopic follow-up using the Micro-Shutter Assembly (MSA \cite{ferruit2022}). Details of the NIRSpec processing are given in ref. \cite{desprez2024}. NIRSpec data have been reduced using the JWST pipeline for stage 1 corrections and then the msaexp \cite{brammer2022} package to create wavelength calibrated, background subtracted 2D spectra. A 1D spectrum is extracted from the 2D using an optimal extraction based on the source spatial profile. The redshift (and its uncertainty) of CANUCS-LRD-z8.6, $z=8.6319\pm 0.0005$, was determined by performing a non-linear least-squares fit simultaneously to the \Hbeta, \OIIIL[4960]\ and \OIIIL[5008]\ emission lines. Each line was modelled with a single Gaussian and the ratio of \OIIIL[4960]\ to \OIIIL[5008] was fixed by atomic physics.

The MACS1149 cluster strong lensing model was derived using \texttt{Lenstool} software \cite{jullo2007} and the catalog of 91 multiple images with spectroscopic redshifts, derived from CANUCS data (Rihtaršič et al., in prep.). As the distance of the LRD from the cluster centre is large ($\sim 3$ arcminutes), the contribution from the cluster lens model alone is small ($\mu=1.060\pm 0.003$). However, the LRD is only 0.54 arcsec away from a foreground galaxy 5112688 at photometric redshift $z_{\rm phot}= 0.49 ^ {+0.05}_{- 0.07}$ and stellar mass $\log (M_*/M_\odot)=  8.0 \pm 0.1$ (derived with \texttt{DenseBasis}). We evaluated the combined magnification in two ways. First, we constructed a \texttt{Lenstool} model containing the cluster model and the foreground galaxy at cluster redshift $z=0.54$ (we verified that using a slightly lower photometric redshift for the galaxy did not affect the results). We modelled the foreground galaxy as a singular isothermal sphere, where the integrated velocity dispersion ($\sigma=40^{+4}_{-2}$  $\mathrm{km s^{-1}}$) was derived from $\log M_*$ and the stellar mass Tully-Fisher relation \cite{mcgaugh2015}. We find that the combined model still yields only a low magnification of $\mu=1.13^{+0.02}_{-0.01}$. Alternatively, we computed the magnification by modelling the galaxy as a dual Pseudo Isothermal  ellipse \cite{eliasottir2007}, following the scaling relations of other cluster members and by using its F160W magnitude of $25.28\pm 0.03$ (we verified that the magnitude uncertainty has a negligible impact on magnification). The scaling relations for parameters were constrained with \texttt{Lenstool} in the inner cluster regions. This method gives a more modest total magnification of $\mu=1.066^{+ 0.004}_{- 0.002}$. We take the latter value as our best magnification estimate $\mu=1.07^{+ 0.08}_{- 0.01}$. Considering that the uncertainty derived from different estimates is small, we correct masses, luminosities, and sizes presented in this work for CANUCS-LRD-z8.6 for a constant $\mu=1.07$. 

Extended Data Figure \ref{fig:phot} shows the photometry of our target in 10 bands.  We fit CANUCS-LRD-z8.6 with \texttt{Galfit}\ \cite{peng2010} and confirm that it is spatially unresolved in all filters (see Extended Data Figure \ref{fig:galfit}). From Galfit modelling, the object is consistent with a point source in all observed NIRCam filters. We perform a more refined fit accounting for the effect of gravitational lensing with \texttt{Lenstruction} \cite{birrer2015, birrer2018,yang2020} to place a more stringent limit on the physical size. \texttt{Lenstruction} performs forward modelling accounting for lensing and the instrumental point spread function. We use lensing maps from the main cluster model, which yields a conservative magnification estimate of $\mu=1.056$, and we choose a clear single star as PSF reference. We use 20 mas image in the F150W filter as this filter comes with the smallest PSF size while still retaining enough flux. The half-light radius of the object results to be smaller than $<0.015$\,arcsec with $95\%$ confidence. This corresponds to an upper limit on the physical half-light radius of $70$\,pc.

\subsection{Continuum and emission line fitting.}
\label{sec:fitting}

The emission lines are fitted to the 1D spectrum of CANUCS-LRD-z8.6 using single or multiple Gaussian components (see below for details). LRDs often have a characteristic continuum shape, with a blue color in the rest-frame UV and red in the rest-frame optical. This is why we split the continuum emission of CANUCS-LRD-z8.6 in two parts modelled by two independent power-laws. In particular, we divide the spectrum in two parts visually setting $\lambda_{\rm rest,sep} = 3400 \AA$, wavelength at which the continuum slope changes sign. Any choice for $\lambda_{\rm rest,sep}$ in the range $3300-3600 \AA$ led to perfectly consistent results. Although our spectrum does not display a prominent break at this location, we note that the wavelength at which our spectrum changes slope is similar to that of breaks observed in other LRDs \cite{wang2024,juodzbalis2024} which have been interpreted as Balmer breaks. We discuss a possible physical interpretation of the spectral shape in Sect. \ref{sec:sed-fitting}. We fit the two parts (UV and optical) of the spectrum separately (see Figure \ref{fig:spectra}). The spectrum is fitted accounting for the well-known variation of prism resolution with wavelength (see below).

No Lyman-$\alpha$ emission has been detected, and the shape of the spectrum around Lyman-$\alpha$ seems to indicate the presence of a damping wing \cite{banados2018,greig2019,totani2006}. The analysis of Lyman-$\alpha$ damping wing is beyond the scope of this paper. Therefore, we exclude the part of the spectrum with $\lambda_{\rm rest} < 1320 \AA$, avoiding any contamination from a possible damping wing given the damping wing's size commonly found in the literature ($\sim 2000-3000$ km s$^{-1}$, see e.g., \cite{greig24,umeda24}). Above $\lambda_{\rm rest} = 1320 \AA$, any detected emission line has been modelled with a single or multiple Gaussian components in case of line blending or broad emission.

The doublet $\OIIIall$ has been fitted fixing the ratio between the peak fluxes (peak$_{\OIIIL[4959]}/{\rm peak}_{\OIIIL[5007]}=0.335$) and the wavelength separation ($\Delta \lambda = 47.94$ \AA) of the two emission lines, and using the same FWHM for both emission lines. Similarly, we fit the $\NeIII$ and $\NeIIIb$ doublet adopting a ratio of 0.301 between the latter and the former, and a rest-frame wavelength separation of $98.73$ \AA, and considering the same FWHM for both emission lines. This reduces to 3 the number of free parameters for both the $\OIII$ and the $\NeIII$ doublets. Given the prism resolution, \NeIII is still blended with \Hzeta, \Heta and \HeIL (see the following section).

Apart from these two doublets, there are other 6 emission lines detected: \NIV, \CIV, \Hdelta, \Hgamma, \OIIIL[4368], and \Hbeta. Each emission line is fitted with a single Gaussian with the only exception of \Hbeta, which shows signatures of a broad emission (see Extended Data Table \ref{tab:line-results}). When trying to fit \Hbeta with one single Gaussian component, the resulting $\sigma_\Hbeta$ is greater than $\sigma_{\OIIIL[5008]}$ by more than $7\%$, which is the expected difference due to the poorer spectral resolution at $\lambda_\Hbeta$. Thus, the \Hbeta emission line is modelled with two Gaussians accounting for both the narrow and broad components.

For the UV part of the spectrum, we have 8 free parameters in total (i.e., peak flux, peak wavelength, and FWHM for \NIV and same for \CIV, power-law exponent and normalization for their underlying continuum), while for the optical part we have 24 free parameters: i.e., peak flux, peak wavelength, and FWHM for \NeIII, \OII, \Hdelta, \Hgamma, \Hbeta$_{\rm narrow}$, \OIIIL[5007]; peak flux and FWHM for \Hbeta$_{\rm broad}$; peak flux and wavelength for \OIIIL[4364], given that we fixed FWHM$_{\OIIIL[4364]}={\rm FWHM}_{\OIII[5008]}$; power-law exponent and normalization for their underlying continuum. We explore the parameter space for each part of the spectrum using a Markov chain Monte Carlo (MCMC) algorithm implemented in the \texttt{EMCEE} package \cite{foreman2013}, assuming uniform priors for the fitting parameters, considering $\sim 5$ walkers per parameter and 2000 trials (the typical burn-in phase is $\sim$ 200 trials). Priors on the FWHM are tight, depending on the resolution of the prism, with the exception of the FWHM$_{\rm broad \Hbeta}$. More precisely, the prior on the FWHM of the narrow component of every fitted emission line is set to be $\rm FWHM^{\rm prior}_{narrow line}\in [1,2]$ spectral resolution elements. The size of the spectral resolution element at the peak wavelength of each fitted line is derived considering the well-known variation of the prism resolution with wavelength \cite{jakobsen+2022}. 

We compute the integrated fluxes by integrating the best-fitting functions for each emission line. In the Extended Data Table \ref{tab:line-results}, we report the fluxes and widths of the fitted emission lines. Unless otherwise stated, we report the median value of the posterior, and $1\sigma$ error bars are the 16th and 84th percentiles. Upper or lower limits are given at $3\sigma$.

\subsection{\Hgamma and \OIIIL[4364].}
 \label{sec:hgamma}

Given the resolution of the prism, \Hgamma is blended with \OIIIL[4364]; nonetheless a clear peak at the nominal \OIIIL[4364] wavelength is observed (see Extended Data Figure \ref{fig:test-Hg}). Therefore, we fitted the blend using two Gaussian components and the results are shown in the Extended Data Table \ref{tab:line-results} (see Fiducial). Since we detected significant broad \Hbeta emission, a broad \Hgamma component could be present along with the narrow \Hgamma and \OIIIL[4364] emission lines. We try to evaluate its impact on our results considering the detection of the broad \Hbeta emission. Hence, we re-fitted the spectrum adding an additional Gaussian component to the \Hgamma-\OIIIL[4364] blend with ${\rm FWHM}_{\Hgamma}={\rm FWHM}_{\Hbeta}$, $\lambda^{\rm peak}_{\rm \Hgamma_{broad}}=\lambda^{\rm peak}_{\rm \Hgamma_{narrow}}$ and the $F^{\rm peak}_{\rm \Hgamma_{broad}}/F^{\rm peak}_{\rm \Hbeta_{broad}}$ ratio corresponding to Case~B recombination (see Test 1 in Extended Data Table \ref{tab:line-results}). Alternatively we also fitted the broad H$_\gamma$ component considering $F^{\rm peak}_{\rm \Hgamma_{narrow}}/F^{\rm peak}_{\rm \Hgamma_{broad}}=F^{\rm peak}_{\rm \Hbeta_{narrow}}/F^{\rm peak}_{\rm \Hbeta_{broad}}$ (see Test 2 in Extended Data Table \ref{tab:line-results}). The spectral resolution and sensitivity of our data do not allow us to be conclusive regarding the presence of a broad \Hgamma component, as shown in Extended Data Figure \ref{fig:test-Hg}. Even though the broad \Hgamma over-predicts the data at $\lambda_{\rm rest}\sim 4300\AA$, this is within $1-2\sigma$, leading to good residuals. Based on the reduced $\chi^2$ criteria, the preferred solution is the one without the broad \Hgamma component (Fiducial), however the other two tests give still reasonably good fits. Therefore, hereafter, we will present primarily the results of our Fiducial fit, and we will discuss the uncertainties introduced by the possible broad \Hgamma component using the results from Tests 1 and 2.

\subsection{Dust correction.}
\label{sec:dust-corr}

In order to estimate the electron temperature, $T_{e}$, and the gas-phase metallicity (hereafter metallicity), O/H, line fluxes need to be corrected for dust reddening. We derive the nebular reddening, $E(B-V)_{\rm neb}$, using the observed ratio of H Balmer lines, \Hbeta and \Hgamma, assuming the Calzetti attenuation law \cite{calzetti1996}. Indeed, the attenuation curve of high-z galaxies is found to be consistent with the Calzetti law. Regarding the observed \Hgamma/\Hbeta ratio, we consider the three cases described in the previous section, depending on whether and how a broad \Hgamma component is included. To derive the reddening, we could have also used \Hdelta but, given the low S/N of this line (lower than for \Hgamma), we cannot evaluate the possible uncertainties introduced by the presence of a broad component. The intrinsic Balmer ratios are computed using \texttt{pyneb} \cite{luridiana2015} assuming $T_e=10^4$ K, and $n_e=10^3 ~\rm cm^{-3}$; results remain in agreement within errorbars even if considering $T_e=2\times 10^4$ K, and $n_e=10^4 ~\rm cm^{-3}$. The derived nebular reddening and dust attenuation are reported in Extended Data Table \ref{tab:ratios}. We note that the negative value found for our Fiducial model suggests the presence of a broad \Hgamma component, which adjusts the dust attenuation to a more reasonable value (see also Sect. \ref{sec:sed-fitting} for a comparison with the stellar $A_V$ from SED fitting). Emission line ratios are then computed using the reddening-corrected fluxes (see Extended Data Table \ref{tab:ratios}). By definition, due to the proximity of the involved lines, \OIIIL[5008]/\Hbeta, \NeIII/\OII[3727], and \CIV/\NIV show almost no dependence on the reddening correction. These are the line ratios of interest for our study.  \OII[3727]/\Hbeta shows a variation of $\sim 0.3$ dex comparing the Fiducial with Test 1/2 models (when considering a broad \Hgamma component).

\subsection{Line blending and contamination correction.}
\label{sec:contam}

When measuring \NeIII, we also include flux from \Heta, \Hzeta ($\lambda=3890.17$~\AA) and \HeIL, both of  which are  blended with \NeIII at the resolution of the prism (hereafter \NeIIIe$_{\rm blend}$ = \NeIII + \Hzeta + \HeIL + \Heta). To estimate the contamination, we proceed as follows. 
For \Hzeta and \Heta, we calculate the dust-corrected ratio \Hdelta/\NeIIIe$_{\rm blend}$ from the fiducial fit (see Extended Data Table \ref{tab:ratios}), finding a value of $0.63^{+0.28}_{-0.20}$. This means that the ratio \Heta/\NeIII$_{\rm blend}$ will be less than 0.18 and the ratio \Hzeta/\NeIIIe$_{\rm blend}$  will be less than 0.25, assuming the Balmer ratios from Case~B recombination,
$T_\mathrm{e}=20,000$~K and $n_\mathrm{e}=10^4~\mathrm{cm}^{-3}$ \cite{storey+hummer1995}. 
For \HeIL, the estimate is unfortunately unfeasible, since we cannot detect the \HeIL[5877] line needed to infer the \HeIL contamination to \NeIII.  Thus we have
 $(\Hzeta + \Heta)/ \NeIIIe_{\rm blend} < 0.43$, implying $\NeIII/\NeIIIe_{\rm blend} >0.57$. This is an upper limit on the contamination because, as we noted, we could not consider the effect of \HeIL. Moreover, considering the dust-corrected fluxes from either Test 1 or Test 2 would imply \Hdelta/\NeIIIe$_{\rm blend}= 0.5 \pm 0.2, 0.55\pm 0.2$, and thus $\NeIII/\NeIIIe_{\rm blend} >0.66, 0.63$, respectively.  In light of these difficulties, we do not apply the estimated correction. However, we show its magnitude for the fiducial case in the relevant figures.

 \subsection{Electron temperature and metallicity.}
 \label{sec:Te}

 We detect the auroral \OIIIL[4364] line, which can be used together with the \OIIIL[5008] to derive the electron temperature and gas-phase metallicity \cite{izotov2006, curti2017, maiolino2019}. Indeed, the electron temperature, $T_e(\OIII)$, of the high-ionization $\rm O^{2+}$ zone of the nebula is computed from the dust-corrected \OIIIL[4364]/\OIIIL[5008] ratio (hereafter RO3). In each of the three cases discussed before, we find a high RO3 possibly indicating the presence of an AGN, a powerful ionizing source. Indeed, in Extended Data Figure \ref{fig:Te} we compare the observed value for RO3 with models from \texttt{pyneb}. For the fiducial dust-corrected RO3, $T_e(\OIII)= 4.0_{-1.2}^{+1.6}\times 10^4$ K, higher than the temperature usually found in normal star-forming galaxies ($T_e(\OIII)\sim 1-2 \times 10^4$ K) \cite{sanders2023, curti2023}. Even when considering the presence of the broad \Hgamma component, $T_e(\OIII)$ is high within the uncertainty, at least $>2\times 10^4$ K. Evidently from Extended Data Figure \ref{fig:Te}, our result is insensitive of the electron density within a range of $n_e=10^2-10^4 \rm ~cm^{-3}$. Using the models of ref. \cite{nicholls2020}, we obtain a consistent result, having $T_e(\OIII)= 3.9_{-1.0}^{+1.6}\times 10^4$ K. Other extreme RO3 have been found in other galaxies at same redshift \cite{katz2023, kokorev2023, mowla2024}, at $z\sim4$ \cite{kokorev2024b}, and are also found in low-z Seyfert galaxies \cite{nagao2001, baskin2005, binette2022}. In particular, similarly to our source, ref. \cite{kokorev2023} found extreme high electron temperature in an LRD at $z=8.5$ also showing broad \Hbeta emission with FWHM$=3439 \rm ~km ~s^{-1}$. These features led the authors to identify the LRD as a broad-line AGN, even in the absence of high ionization UV emission/absorption lines (e.g. \CIV, \NIV, \NV).
 Moreover, such a high ratio of \OIIIL[4364]/\Hgamma as ours ($\log(\OIIIL[4364]/\Hgamma)\sim -0.3$) is observed in AGN \cite{perna2017,brinchmann2023}. This can be further seen by comparing nearby AGN and star-forming galaxies from SDSS \cite{abazajian2009} in the\OIIIL[4364]/\Hgamma versus \OIIIL[5008]/\OIIIL[4364] line ratio diagram as done in ref. \cite{ubler2024} for ZS7. In this diagram, local star-forming galaxies and AGN separate into two parallel sequences, with AGN occupying a region of higher electron temperatures and having elevated ratios of \OIIIL[4364]/\Hgamma. Indeed, our target lies at the extreme end of the AGN population, the farthest from the star-forming galaxies. 

 We want to derive a first-order estimate of the metallicity of our source from the total oxygen abundance $\rm O/H = O^{2+}/H^+ + O^+/H^+$. To compute $\rm O^+/H^+$, $T_e(\OII)$ is required. However, none of the \OII transitions (\OII[3727], \OIIalllow) is or can be detected in the spectra. Therefore, we use the relation of ref. \cite{campbell1986}:

 \begin{equation}
     T_e(\OII) = 0.7 \times T_e(\OIII)+3000 ~\rm K
    \label{eq:TeO2}
 \end{equation}

 \noindent For our fiducial case, $T_e(\OII)=3.1\times 10^4$ K. Ionic and total oxygen abundances are computed using \texttt{pyneb}, assuming that all O is in either the $\rm O^{2+}$ or $\rm O^+$ states inside HII regions. Indeed, $\rm O^{3+}$ may be neglected considering that it is $<5\%$ of the total O even in extremely high-ionization systems \cite{berg2018,berg2021} and it is negligible given the uncertainty of our computations. The $\rm O^{2+}/H^+$ ratio is derived from the dust-corrected \OIII[5008]/\Hbeta, the $\rm O^{+}/H^+$ ratio from the dust-corrected \OII[3727]/\Hbeta upper limit, and we assume the $T_e(\OIII), T_e(\OII)$ derived above for the fiducial case. Given that the upper limit on \OII[3727]/\Hbeta ratio and that the highest allowed electron temperature in Pyneb models is $T_e=3\times 10^4$ K, we can derive an upper limit on the metallicity. Indeed, $O/H$ ratios decrease at fixed line ratios and increasing electron temperature. The inferred metallicity of our source is $12+\log(\rm O/H)<7.9$ or $\log(Z/Z_\odot)<-0.7$. The upper limit becomes more stringent if considering the presence of the broad \Hgamma component (either Test 1 or 2), having $12+\log(\rm O/H)<6.9$ or $\log(Z/Z_\odot)<-1.8$. As a word of caution, we mention that the possible presence of very high-density regions ($\log(n_e)>>4$) have an impact on the observed flux of the \OIIIL[5008] line due to the collisional de-excitation of the lower level \OIIIL[5008] bearing transition \cite{marconi2024, ji2024}. However, the available data prevent us from quantifying this effect since the density distribution cannot be derived.
 
 An additional evidence of CANUCS-LRD-z8.6 being metal-poor is given by the results obtained from the comparison of the `OHNO' diagnostic, which relates the \OIII[5008]/\Hbeta with \NeIII/\OII[3727], with photo-ionization models (see Figure \ref{fig:diagnostic}). This diagnostic has been used to distinguish between the star-forming or AGN nature of galaxies at low- and high-z \cite{backhaus2022, cleri2022, larson2023}. In Figure \ref{fig:diagnostic} we compare our results with $z\sim 0$ SDSS AGNs (galaxies) as blue (pink) colormap with contours, $z\sim 2$ MOSDEF galaxies and AGNs (black contours), SMACS~06355, 10612 and 04590 (red diamonds \cite{trussler2023}; the left-most square of the three is 06355, the type-II AGN identified by \cite{brinchmann2023}), the type-I AGN host GS~3073 at $z=5.55$ (filled pink and hollow diamond, the latter estimating the flux of \NeIII based on the Case B assumption modulated by the median dust attenuation; \cite{ji2024}), and the $z=7.15$ AGN type-I's host galaxy ZS7 (yellow cross and diamond, depending on whether line fluxes are computed from the BLR location or \OIII centroid, respectively; \cite{ubler2024}). Overlaid are the AGN photoionization models of ref. \cite{feltre2016} at hydrogen densities $\log n{\rm [cm^{-3}]}=2.0$ (gray scale) and $\log n{\rm [cm^{-3}]}=4.0$ (coloured scale). Our observed `OHNO' ratios point towards $\log(Z/Z_\odot)<-1.0$.  Our stringent lower limit on \NeIII/\OII[3727] also indicates high-ionization (ionizing parameter of $\log U\sim -1.5$), and similar results have been reported in other AGN candidates at $z> 8$ \cite{larson2023, kokorev2023}. Alternatively, we also compute the gas-phase metallicity just from the fiducial dust-corrected \OIII[5008]/\Hbeta ratio using the empirical relation found in refs. \cite{sanders2023, curti2020, nakajima2023}, assuming the narrow lines to be dominated by star formation. This implies a low metallicity of  $12+\log(\rm O/H)=7.08_{-0.12}^{+0.14}$ ($Z\sim 0.02 ~Z_\odot$), $12+\log(\rm O/H)=7.40_{-0.11}^{+0.13}$ ($Z\sim 0.05 ~Z_\odot$), and $12+\log(\rm O/H)=7.28_{-0.12}^{+0.15}$ ($Z\sim 0.04 ~Z_\odot$), respectively, which are consistent with the results from the `OHNO' diagnostic indicating $Z<<0.1 Z_\odot$.

 \subsection{\CIV, \NIV, \NV, and \NeV.}
 \label{sec:other-lines}

High-ionization lines requiring photoionization energy $>50-60$ eV, such as \NIV, \NV, and \NeV are signatures of the presence of an AGN. Even though \CIV is also usually associated with the presence of a central AGN \cite{maiolino2024, mazzucchelli2023}, it is not an unambiguous tracer of an AGN in the absence of other signatures (e.g., \NIV, broad emission), since it has also been detected in some low-mass low-Z galaxies at high-z. We have clear evidence of \NIV and \CIV emission, while both \NV and \NeV remain undetected, as well as \CIII. We note, however, that the resolution of the prism does not allow us to assess the presence of \NV since it is blended with the Ly$\alpha$ and its damping wing. In many AGNs some of these emission lines are either very weak or undetected if the S/N is not high enough \cite{kuraszkiewicz2004, nagao2006, cleri2023}. For instance, in the type 1.8 AGN GS-3073 at $z=5.5$ \cite{grazian2020, ubler2023, vanzella2010}, the \NV is five times weaker than \NIV, which would be totally undetected in our spectrum. Similarly, \NV is undetected in GNz-11 \cite{maiolino2024} and in other type 1 quasars \cite{glikman2007}, while \NIV is strong. As discussed in \cite{maiolino2024, glikman2007}, \NeV/\NeIII can be quite low in AGNs, down to $10^{-2}-10^{-4}$. Recently, the simultaneous detection of both \CIV and \NIV in galaxies at $z\sim 7$ has been attributed to ionization by dense clusters of massive stars formed during an intense burst of star formation \cite{topping2024}. This interpretation is supported by high observed specific star formation rates (sSFR$>300-1000$ Gyr$^{-1}$) and large \Hbeta equivalent widths (EW$>400-600\AA$). However, for CANUCS-LRD-z8.6, this scenario is unlikely due to its very low inferred sSFR (sSFR$<10$ Gyr$^{-1}$; see Sect. \ref{sec:sed-fitting} and Figure 18 in ref. \cite{topping2024}), indicating that sources other than massive stars are needed to account for its strong ionization.

Given the uncertainties on the dust correction given by the possible presence of a broad \Hgamma component and the absence of the \OIIIlow emission line in the UV, we will just discuss the \CIV/\NIV ratio, which is reddening insensitive, leaving aside the discussion about the C/O or N/O ratios. Assuming that all the nitrogen is in $\rm N^{3+}$, emitted in \NIV, and all the carbon in $\rm C^{3+}$, emitted in \CIV, we obtain a low \CIV/\NIV ratio having dust corrected $\log(\CIV/\NIV)= 0.07 \pm 0.3$.
Assuming a temperature of $40,000$~K as derived from from the \OIIIL[4364]/\OIIIL[5008] ratio and a density of $n_e=10^3$~cm$^{-3}$, we infer a carbon-over-nitrogen abundance of log(C/N)$=-0.75 \substack{+0.05 \\ -0.04}$. Such low C/N abundance ratio is similar to what was reported for some nitrogen-enriched galaxies recently observed at high redshift \cite{isobe2023}, and aligns with abundance patterns measured for dwarf stars in local globular clusters \cite{dorazi2010}, possibly suggesting that material-enriched through the CNO cycle has been effectively ejected via powerful stellar winds from the outermost layers of massive stars \cite{maeder2015,charbonnel2023,watanabe2024}.

 \subsection{Black hole mass and bolometric luminosity of the AGN.}
 \label{sec:bh_prop}

 Robust estimates of BH masses usually come from reverberation mapping studies, which unfortunately are not feasible at high-z. Therefore, the so-called single-epoch virial mass estimate of $M_{\rm BH}$ is often used \cite{maiolino2023,mazzucchelli2023}, assuming that virial relations are still valid at high-z and considering the continuum or line luminosity and the FWHM of the broad emission lines. For this work, we use the empirically derived relation:

 \begin{equation}
    \dfrac{M_{\rm BH}}{\rm M_\odot}=\alpha  \biggl(\dfrac{L_\lambda}{10^{44}\rm erg ~s^{-1}}\biggr)^\beta \biggl(\dfrac{\rm FWHM_{line}}{10^3\ \rm km ~s^{-1}}\biggr)^2,
\end{equation}

\noindent where the best-fit values for the scaling parameters $\alpha, \beta$ depend on the respective emission lines and/or monochromatic luminosity $L_\lambda$ chosen. For instance, considering the H$\beta$ line one has $\alpha=(4.4\pm 0.2)\times 10^6, \beta=0.64\pm 0.2$ at $L_\lambda=L_\Hbeta$ or, considering the continuum luminosity at rest-frame 5100 \AA, $L_{5100\AA}$, it is found $\alpha=(4.7\pm 0.3)\times 10^6, \beta=0.63\pm 0.06$ at $L_\lambda=\lambda L_{5100 \AA}$ \cite{greene2005}. The BH masses derived from these relations can be found in Extended Data Table \ref{tab:bh-prop}. Alternatively, we also used the relations of \cite{vestergaard2009}, finding a systematic rise in BH mass of $\sim 0.15-0.2$ dex. These relationships are calibrated to the most recently updated and robust mass determinations from reverberation mapping. The majority of reverberation mapping studies have been conducted using \Hbeta on low redshift AGN \cite{bentz2013,barth2015,grier2017,malik2023}. For high-z sources, the MgII or \CIV line is often utilised. However, this involves applying additional scaling from the H$\beta$ line to formulate the virial mass based on other lines \cite{shen2011}. These relations have been used to measure BH masses for thousands of sources with an estimated uncertainty of about factor 2-3 (i.e. dex=0.3-0.5 \cite{shen2013}), when using either \Hbeta or MgII. Estimates based on the high ionization \CIV line are even more uncertain ($>0.5$ dex), as this line shows large velocity offsets, implying significant non-virialized motions \cite{mejia2018, park2017}. Moreover, there is mounting evidence that large CIV blueshifts ($>2000$ km s$^{-1}$) are more common at $z>6$ than at lower redshifts \cite{meyer2019, schindler2020,matthews2023}. Since the prism resolution of our data does not allow us to distinguish between the narrow and broad \CIV component, either from the BLR or from outflows, we do not use the detected \CIV emission line to infer the $M_{\rm BH}$. We report our estimate for the BH mass of CANUCS-LRD-z8.6 from both \Hbeta and $L_{5100\AA}$ in Extended Data Table \ref{tab:bh-prop} for our fiducial fit.

\noindent From the BH mass measurements ($M_{\rm BH, \Hbeta}, M_{\rm BH, 5100 \AA}$), we calculate the Eddington luminosity:

\begin{equation}
    L_{\rm Edd, \Hbeta/ 5100\AA} = 1.3\times 10^{38} \biggl(\dfrac{M_{\rm BH, \Hbeta/ 5100\AA}}{M_\odot}\biggr) {~\rm erg ~s^{-1}}
\end{equation}

\noindent We also compute the bolometric luminosity ($L_{\rm bol}$) of the AGN using the continuum luminosity at 3000 \AA and using the bolometric correction presented by ref. \cite{richards2009}. From the $L_{\rm Edd}$ and $L_{\rm bol}$, we derive the corresponding Eddington ratios $\lambda_{\rm Edd}=L_{\rm bol}/L_{\rm Edd}=0.1$. We report all our results in Extended Data Table \ref{tab:bh-prop}. We find comparable quantities (within 0.2-0.5 dex) also for the Test 1 and Test 2 cases. 

\subsection{Spectral energy distribution fitting}
\label{sec:sed-fitting}

We perform a spectro-photometric fit to the NIRCam photometry and NIRSpec spectroscopy using \texttt{Bagpipes} \cite{carnall2018, carnall2019} with the primary goal of determining the stellar mass for CANUCS-LRD-z8.6. There was no need to scale the spectrum to the photometry. 
In the \texttt{Bagpipes} SED fitting procedure, we fix the redshift to the spectroscopic redshift of 8.63, and we assume a double power law (DPL) star formation history (SFH), Calzetti dust attenuation curve \cite{calzetti2000}, and Chabrier initial mass function (IMF) \cite{chabrier2003}. The priors for the fitting parameters are reported in Extended Data Table \ref{tab:res-bagpipes}. We fixed the ionization parameter to $\log(U)=-1.5$, which is derived from the `OHNO' diagnostic (see Sect. \ref{sec:Te}). We also set the range of metallicity considering the highest upper limit derived from observations, $Z<0.2~Z_\odot$ (see Sect. \ref{sec:Te}). We checked that our results do not change when increasing the upper bound of the metallicity range up to $Z = 2.5~Z_\odot$.

We adopt a Calzetti attenuation curve in the SED fitting procedure as the dust attenuation curves of high-redshift galaxies ($z > 6$) are generally found to be flat and lack a prominent UV bump feature \cite{markov2023, markov2024}. In addition to using the Calzetti standard template as our fiducial dust attenuation model, we try to fit the data with an SMC template. Moreover, we adopt a flexible analytical attenuation model \cite{markov2023, markov2024} to better constrain the shape of the dust attenuation curve for our object. The resulting inferred attenuation curve is Calzetti-like, though slightly shallower in the rest-frame UV. We also found that the assumed shape of the dust attenuati
on curve significantly impacts the inferred V-band dust attenuation ($\Delta A_V \sim 0.6 ~\rm{dex}$), which in turn affects fundamental galaxy properties to a lesser extent (e.g., $M_*$, SFR, and stellar age by $0.2-0.4$  dex) due to degeneracies. This is consistent with previous studies which conducted similar analysis \cite{reddy2015,salim2016,topping2022,markov2023}.

Alongside the DPL model, which we use as our fiducial SFH model, we also perform fits with other SFHs, including the non-parametric SFHs from ref. \cite{iyer2019} 
and the Leja model with a continuity prior \cite{leja2019}, and the parametric exponentially declining SFH. We found similar results within uncertainties regardless of the SFH model choice. However, this may be an exception rather than the rule, as some studies in the literature indicate that SFH model selection can significantly impact the inferred galaxy properties \cite{leja2019, topping2020, topping2022, whitler2023, markov2023}.    

Firstly, we fitted the observed SED without including an AGN contribution (no-AGN run). Therefore, to allow reliable estimates of the inferred host galaxy's properties, we subtracted from the observed spectrum the broad \Hbeta component, which is a clear AGN signature, using the best-fitting model shown in Figure \ref{fig:spectra}. We checked that subtracting the \CIV and \NIV did not change the fitting results. We did not treat the UV continuum, since the real AGN contribution in LRDs to the UV flux is still unknown, and we wanted to understand what the properties of CANUCS-LRD-z8.6 would be if all the observed UV light came from stars. Extended Data Figure \ref{fig:sed-fitting} shows the \texttt{Bagpipes} spectro-photometric fit in orange along with the posterior distribution of some quantities of interest and the resulting SFH. Results for the fitting parameters are reported in Extended Data Table \ref{tab:res-bagpipes}. The best-fitting \texttt{Bagpipes} model is able to reproduce most of the shape of the observed spectrum of CANUCS-LRD-z8.6. However, it does not capture some features that can be ascribed to the presence of a powerful ionizing source: (i) the non-detection of \OII[3727] emission while a very bright \OIIIL[5008] emission; (ii) the full \Hgamma-\OIIIL[4364] flux, possibly due to the presence of a broad \Hgamma component; (iii) the red slope of the continuum in the optical regime; (iv) the \CIV and \NIV emissions. Indeed, in this run of \texttt{Bagpipes}, the main excitement mechanism for emission lines comes from stars thus, in our case, simple stellar population (SSP) models cannot reproduce all the observed spectral features.

Consequently, we run \texttt{Bagpipes} including a model
for AGN continuum, broad \Hbeta and \Hgamma emission \cite{carnall2023} (AGN run). In \texttt{Bagpipes}, following ref. \cite{vandenberk2001}, the AGN continuum emission is modelled with a broken power law, with two spectral indices and a break at $\lambda_{\rm rest}= 5100 \AA$. The broad \Hbeta is modelled as a Gaussian varying normalization and velocity dispersion. The broad \Hgamma has the same parameters of the broad \Hbeta but with the normalization divided by the standard ratio from Case B recombination \cite{osterbrock2006}. Extended Data Figure \ref{fig:sed-fitting-AGN} shows the \texttt{Bagpipes} spectro-photometric fit in orange along with the posterior distribution of some quantities of interest and the resulting SFH. The red continuum in the optical is now captured by the best-fitting model, as well as the broad emissions. Furthermore, the \OII emission is dimmer than in the previous run, yet it still does not align with the observed non-detection. The metallicity in both runs (w/o and with AGN) is in agreement with the observed data. The dust attenuation is $\sim 3.5$ times higher than in the previous run, causing the stellar mass to increase by 0.3 dex. We did not set a tight prior on $A_V$ since the observed value is very uncertain (see Sec. \ref{sec:dust-corr} and Extended Data Table \ref{tab:ratios}); indeed it is in agreement within errors with the results of both no-AGN and AGN runs, considering that $A_V=0.44 A_V^{\rm neb}$ assuming a Calzetti dust law \cite{calzetti2000}. However, in order to compare with the results from the no-AGN run, we run \texttt{Bagpipes} including the AGN model as before and setting a tight Gaussian prior around the value of $A_V$ determined from the no-AGN run (AGN-tight run), and we obtained $A_V= 0.7\pm 0.2$ and $\log(M_*/M_\odot)=9.45\pm 0.07$. Discussing the result for the spectral index in the UV regime, $\alpha_\lambda$, our best-fit model favours a very low $\alpha_\lambda$, at the edge of the prior close to $-2$. The continuum slope in the UV is usually found to be within the $(-2,2)$ range of values \cite{vandenberk2001,groves2004a,kewley2006,yue2012}, and $\alpha_\lambda=-2$ gives a good result in terms of residuals, comparable to the no-AGN case. However, we performed an addition run with \texttt{Bagpipes} extending the lower range of the prior on $\alpha_\lambda$ down to $-4$, in order to ascertain the implications on the derived properties. In this case, the best fit $\alpha_\lambda$ is equal to $-2.4\pm 0.1$, and the stellar mass show a slight increase to $\log(M_*/M_\odot)= 10.1\pm 0.1$, while the other properties still remain consistent within errorbars. Finally, the degeneracy between the AGN model, $A_V$ and $M_*$ is evident, and prevents us from obtaining a precise determination of $M_*$. For the aim of this work, we considered the $M_*$ derived from the AGN-run as fiducial, and its error accounts for the uncertainties due to the variation of the SED model, i.e. $\log(M_*/M_\odot)=9.85^{+0.32}_{-0.48}$ (corrected for magnification).  

With these caveats, we hypothesize a physical model of the CANUCS-LRD-z8.6 system which can account for all of the observations and the derived properties. As shown in Extended Data Figure \ref{fig:sed-fitting-AGN}, the AGN dominates the UV continuum when it is included in the fit. Combined with the observed broad \Hbeta and high-excitation UV lines, this suggests our sight-line to the AGN is not heavily dust-obscured. On the other hand, the fit prefers a lower AGN contribution in the rest-optical regime, instead accounting for the red continuum with more dust obscuration of the stellar component. Given the very compact size of the source and the significant SFR inferred from the fit ($\sim 50 ~\rm M_\odot yr^{-1})$, it is reasonable to infer that the bulk of the stellar light is embedded in stellar birth clouds, leading to high obscuration. Ref. \cite{casey2024} show that even small amounts of dust can cause significant obscuration in LRDs given their compactness. Altogether, this points to a highly compact system undergoing an episode of star formation with a high dust covering fraction in which a highly energetic AGN has cleared a sight-line in our direction \cite{nenkova2008,honig2016,garcia2021}. This perhaps points toward CANUCS-LRD-z8.6 being a more evolved system than some other observed LRDs that exhibit less massive BHs and host galaxies, on its way to becoming a system resembling the brightest quasars at $z=6$ rather than the lower luminosity AGNs recently discovered by JWST in this redshift range (see also Sect. \ref{sec:comp-simul}). Extended Data Figure \ref{fig:cartoon} shows a visual representation of this physical configuration.

Even though the run including the AGN component better reproduces the observed data, higher wavelength observations are needed to constrain the real AGN contribution to the observed multiwavelength light of CANUCS-LRD-z8.6. How best to incorporate AGN components in SED fitting for LRDs remains a topic of ongoing debate due to the inability of current data to meaningfully distinguish between different models \cite{barro2024, casey2024, wang2024}. In light of these uncertainties, we decided to use the stellar mass of the AGN-run and to account for the variation arising from the other models (no-AGN, AGN-tight) in the error bars.

\subsection{Comparison with simulations and semi-analytical models}
\label{sec:comp-simul}

In this section, we investigate the possible formation channels for the massive BH powering CANUCS-LRD-z8.6 by comparing the inferred BH mass with predictions from semi-analytical models (SAM) and numerical simulations.

To get a first approximate idea about the possible growth history of the CANUCS-LRD-z8.6's BH, we first assume that this BH has been accreting for its entire history at a fixed rate, expressed as a fraction of the Eddington rate, with a constant radiative efficiency $\epsilon=0.1$. As shown in Figure \ref{fig:bh_growth}, fixing the accretion rate to the observed value ($\lambda_{\rm Edd}=0.1$) requires an extremely heavy BH mass ($M_{\rm seed} > 3\times10^7~M_\odot$) at redshifts higher than 25. This seed mass is higher than any value predicted by theoretical models \cite{Volonteri2021}. This implies that, at earlier epochs, the BH powering CANUCS-LRD-z8.6 must have been accreted at rates higher than the one observed at $z=8.63$. 

Assuming $\lambda_{\rm Edd}=1$ leads to a seed mass of $M_{\rm seed} \sim 10^4~M_\odot$ at $z\sim 25$ or $M_{\rm seed} \gtrsim 10^5~M_\odot$ at $z \sim 15$. This growth path is consistent both with intermediate-mass BHs formed in dense star clusters \cite{PortegiesZwart2004} and with heavy seeds predicted by the direct-collapse BH scenario \cite{Bromm03}.

Assuming $\lambda_{\rm Edd}=1.5$, the required seed mass would be consistent with low-mass seeds (10-100~$M_\odot$) from Pop III stellar remnants at $z\gtrsim 20$ \cite{Madau2001}. This argument suggests that the BH in CANUCS-LRD-z8.6 originates from heavy seeds, constantly growing at a pace close to Eddington, or from light seeds constantly growing at super-Eddington rates.

In the following, we discuss the formation channel of CANUCS-LRD-z8.6's BH more accurately by performing a comparison with SAM predictions. Initially, we consider the results by ref. \cite{Cammelli24}, hereafter C24, which are able to reproduce several important global properties of galaxies both at high and low redshifts, such as the luminosity and stellar mass functions of galaxies from the local Universe up to $z\sim 9$ \cite[][; Cammelli et al, 2025, in prep]{Cammelli24}, and the local $M_{\rm BH}-M_*$ relation. The C24 models are based on the GAEA SAM \cite[e.g.,][]{hirschmann2016, fontanot2020} run on merger trees extracted by using the \texttt{PINOCCHIO} code \cite{munari2017}, 
and we consider here two different seeding models: (i) Pop III.1, a scheme that allows for an early formation of massive seeds ($\sim10^5\ M_{\odot}$) at $z\sim25$ from the collapse of Pop III protostars \cite{banik2019}; this formation mechanism is physically motivated and does not depend on the mass resolution of the simulation; (ii) All Light Seed (ALS), a model that results into seeds of 10-100 $M_\odot$.

For the scope of this work, we compare CANUCS-LRD-z8.6 with the most massive BHs predicted by the C24 models at $z\sim 8$ for the two seeding prescriptions, as shown in the left panel of Exteded Data Figure \ref{fig:models}. We find that all these models fail in capturing the CANUCS-LRD-z8.6's BH mass at $z=8.6$ by several orders of magnitudes. This suggests that exists a population of high-redshift BHs, with CANUCS-LRD-z8.6 representing the most extreme example at $z\sim8.6$, that cannot be explained by these SAMs employing standard prescriptions, although the same models successfully describe larger galaxy and AGN populations from the local up to the high-$z$ Universe.

We further consider the results by ref. \cite{Schneider2023}, which investigates the formation of massive BHs at $z>7$ by means of the SAM Cosmic Archaeology Tool (CAT) \cite{Trinca2022}. 
In this work, the seeding prescription accounts for both light and heavy seeds, and the BH growth can occur in the Eddington-limited (EL) and super-Eddington (SE) regimes. As can be seen from the left panel of Extended Data Figure \ref{fig:models}, in the CAT framework, super-Eddington accretion is essential to assemble a large amount of mass within 500 Myr (see also \cite{Pezzulli2016, Pezzulli2017}) and therefore to reproduce the CANUCS-LRD-z8.6 inferred mass. The EL model predicts BH masses that are consistent with GNz-11 at $z\sim 10$ and CEERS-1019 at $z\sim 8.7$ \cite{larson2023}, but do not exceed $\sim10^7\ M_{\odot}$, therefore being inconsistent with our new data. From this analysis, CANUCS-LRD-z8.6 emerges as a very extreme source at these early epochs.

The main caveat of SAMs is that they cannot fully capture the complex and non-linear interplay between BH accretion and feedback processes. Therefore the growth predicted in SAMs during the SE phase might be too efficient, if compared to more sophisticated models, e.g. hydrodynamical numerical simulations. This is clearly shown by the results of ref. \cite{zhu2022, Sassano2023} that are based on numerical simulations with light seeds growing at super-Eddington pace. From the ref. \cite{zhu2022} (\cite{Sassano2023}) results, it can be seen that early light seeds, even if accreting at super-Eddington rate, can reach a maximum mass of $10^5~M_\odot$ ($10^7~M_\odot$) at $z\sim 6$, thus being unable to reproduce the BH masses estimated so far at $z>6$. For this reason, most of the numerical hydro-dynamical simulations of BH formation and growth\footnote{The accretion rate onto the BHs is modeled according to the Bondi-Hoyle-Lyttleton prescription \cite{Hoyle1939, Bondi1944, Bondi1952}, with a boost factor $\alpha$ used as a correction factor for the spatial resolution of the gas distribution surrounding the BH.} assume a heavy seed prescription ($M_{\rm seed}>10^5~M_\odot$) to reproduce the large masses of BHs powering $z\sim 6$ quasars. 

In what follows, we separately discuss predictions from simulations that cap the BH accretion to the Eddington limit and those that allow for super-Eddington growth. In the middle panel of Extended Data Figure \ref{fig:models}, we report the results of EL simulations. We find that the only simulation that can reproduce the CANUCS-LRD-z8.6 BH mass is the reference run by \cite{zhu2022}, whereas ref. \cite{bennett2024}, when using the numerical recipe of the \texttt{FABLE} suite \cite{henden2018}, and ref. \cite{Bhowmick2022} predict a BH mass that is $\sim$ 1 order of magnitude smaller. Interestingly, all the simulations reported in this panel, though being inconsistent with CANUCS-LRD-z8.6, are capable of reproducing AGN candidates such as GNz-11 at $z\sim 10$ and CEERS-1019 at $z\sim 8.7$ \cite{larson2023}, and the estimated masses of BHs powering $z\sim 6$ quasars, apart from the most extreme case of J0100+2802 ($M_{\rm BH} \sim 10^{10}~M_\odot$, \cite{mazzucchelli2023}). This underlines once more that the BH mass of CANUCS-LRD-z8.6 is the most challenging for theoretical models. We further notice that the simulations by \cite{zhu2022} predict a BH mass at $z\sim 8$ that is $\sim 2$ orders of magnitude smaller with respect to the reference run if a radiative efficiency larger than only a factor of two is considered. This clearly shows how sensitive predictions from numerical simulations are to the feedback prescriptions implemented. 

We now move to numerical simulations of heavy BH seeds' growth, including super-Eddington accretion, shown in the right panel of Extended Data Figure ~\ref{fig:models}. First of all, we note that the reference run by \cite{bennett2024} is able to reproduce not only CANUCS-LRD-z8.6 but also $z\sim 6$ quasars, including the extreme case of J0100+2802. With respect to the original recipe employed in the \texttt{FABLE} suite (shown in the middle panel), in the reference run the authors apply the following variations: (i) reduce the halo mass where BH seeds are placed (from $M_h = 5 \times 10^{10}~h^{-1}~\rm M_\odot$ to $M_h = \times 10^9~h^{-1}~\rm M_\odot$), effectively resulting in earlier BH seeding (from $z\sim 13$ to $z\sim 18$); (ii) reduce the overall AGN feedback; (iii) allow for mild super-Eddington accretion ($\lambda_{\rm Edd}=2$). All these changes promote early BH growth, which emerges as a necessary condition to explain the BH mass of early AGN and quasars. 
Interestingly, this simulation supports a scenario in which CANUCS-LRD-z8.6 represents a progenitor of the most massive QSOs at $z>6$, such as J0100+2802. 

We further notice that the run \textit{Bh22d} of ref. \cite{Bhowmick2022} (see also the similar setup used in ref. \cite{Ni2022}) predicts a rapid mass assembly consistent with CANUCS-LRD-z8.6, if the accretion rate is boosted ($\lambda_{\rm Edd}=2$, $\alpha=100$) \textit{and} the radiative efficiency is low ($\epsilon=0.1$), which in turn lowers the AGN feedback effect on the BH growth. Similarly, \cite{zhu2022,Valentini2021} find the AGN feedback to be the most limiting factor in BH growth.

Notably, in \cite{zhu2022}, the BH  grows less in the super-Eddington regime due to the excessive feedback. However, this conclusion is sensitive to the detailed numerical implementation. Ref. \cite{Lupi2024} explored the BH growth with high-resolution numerical simulations with a comprehensive model of AGN feedback in the super-Eddington regime. They find that the jet power in the super-Eddington regime is a critical factor in regulating the accretion rate onto the BH, because of its ability to remove the fueling gas, as also found in other works \cite{massonneau2023}. Their run with low feedback predicts a BH mass at $z\sim 8.5$ only a factor of $\sim 2$ smaller than CANUCS-LRD-z8.6.

The comparison among results from different numerical simulations emphasizes how complex is the modeling of BH seeding, accretion rate, and AGN feedback, and how important it is to collect observational data as the one provided in this work. CANUCS-LRD-z8.6 poses significant challenges to both hydrodynamical simulations and semi-analytical models. Its existence requires a rapid and efficient assembling of $10^8~M_\odot$ in only 500~Myr, thus providing stringent constraints to seeding prescriptions, feedback recipes, and accretion models in theoretical models.



\printbibliography[segment=2]

\end{refsegment}

\clearpage
\setcounter{figure}{0}   
\setcounter{table}{0} 
\captionsetup[table]{name=Extended Data Table}
\captionsetup[figure]{name=Extended Data Figure}

\section*{EXTENDED DATA}\label{sec_ED}

\begin{figure}[h]
    \centering
    \includegraphics[width=0.9\linewidth]{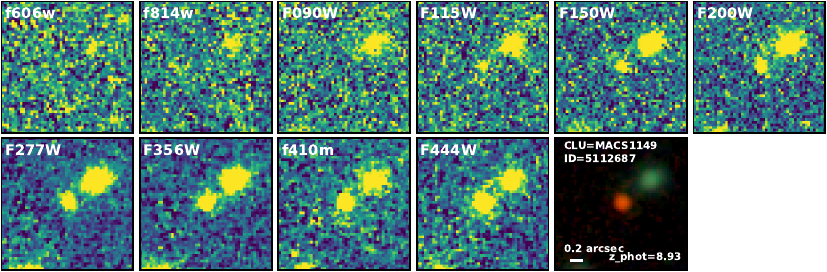}
    \caption[]{Photometric data in 10 filters (f435w, F606W, F814W, F090W, F115W, F150W, F200W, F277W, F358W, f410m, F444W) and RGB image (bottom right panel) of 5112687. The RGB image is made combining psf-matched photometry at F090W, F200W, F444W.}
    \label{fig:phot}
    \vspace{-0.5cm}
\end{figure}

\begin{figure}
    \centering
    \includegraphics[width=0.65\linewidth]{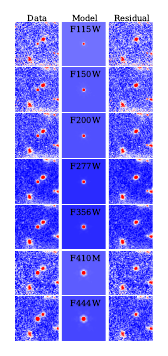}
    \caption[]{Results of modeling CANUCS-LRD-z8.6 as a single point source with GALFIT. The observed image (left), point source model (center), and residual (right) are shown for each filter as indicated.}
    \label{fig:galfit}
\end{figure}

\begin{figure}
    \centering
    \includegraphics[width=0.32\linewidth]{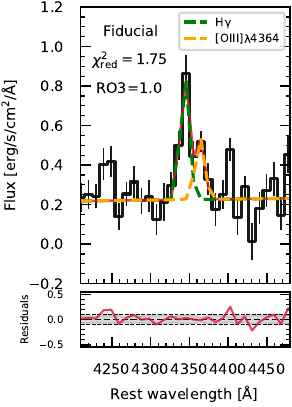}
    \includegraphics[width=0.32\linewidth]{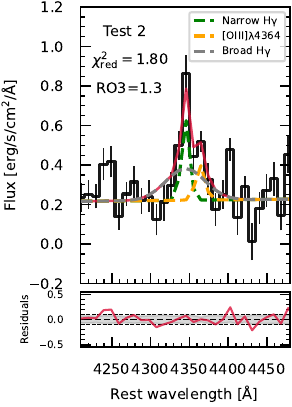}
    \includegraphics[width=0.32\linewidth]{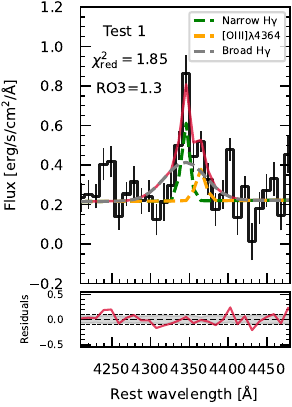}
    \caption[]{Zoom-in of the \Hgamma, \OIIIL[4364] emission lines from the 1D spectrum of 5112687. Top panels: the red solid line is the total best-fitting curve. The value of the reduced $\chi^2$ of the fit is reported in the top left corner of each panel, as well as the ratio between \OIIIL[5008] and \OIIIL[4364] (RO3, in dex). Left: fiducial fit (see also Figure \ref{fig:spectra}) considering only a narrow component for \Hgamma (green dashed line), and \OIIIL[4364] (yellow dashed line). 
    Centre: test 2 fit adding a broad \Hgamma component (gray dashed line) and considering the observed ratio between the narrow and broad \Hbeta component to scale the broad \Hgamma. Right: test 1 fit adding a broad \Hgamma component (gray dashed line) and considering case B recombination ratios between the broad \Hgamma and the broad \Hbeta to scale the broad \Hgamma.  Bottom panels: residuals of the fits shown in the top panels (red solid line). The black dashed lines mark the 0 level and the average $\pm 1\sigma$ noise level. The gray shaded area is the noise.}
    \label{fig:test-Hg}
\end{figure}

\begin{figure}
    \centering
    \includegraphics[width=0.9\linewidth]{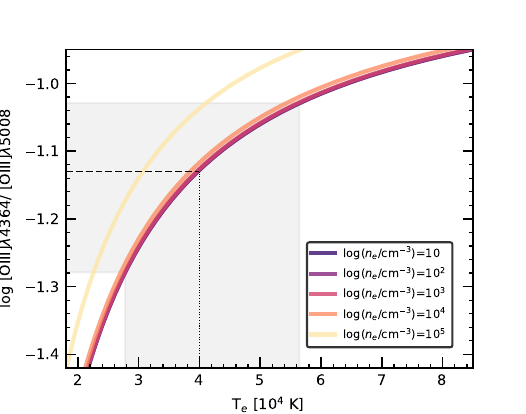}
    \caption[]{Variation of the the \OIIIL[4364]/\OIIIL[5008] ratio, as a function of the electron temperature, $T_e(\OIII)$, for different electron densities, $n_e=10-10^5 \rm ~cm^{-3}$ (as specified in the legend). The models have been generated with \texttt{pyneb}. The horizontal dashed line marks the observed fiducial dust-corrected \OIIIL[4364]/\OIIIL[5008] ratio, and the gray shaded region is its error.}
    \label{fig:Te}
\end{figure}

\begin{figure}
    \centering
    \includegraphics[width=1\linewidth]{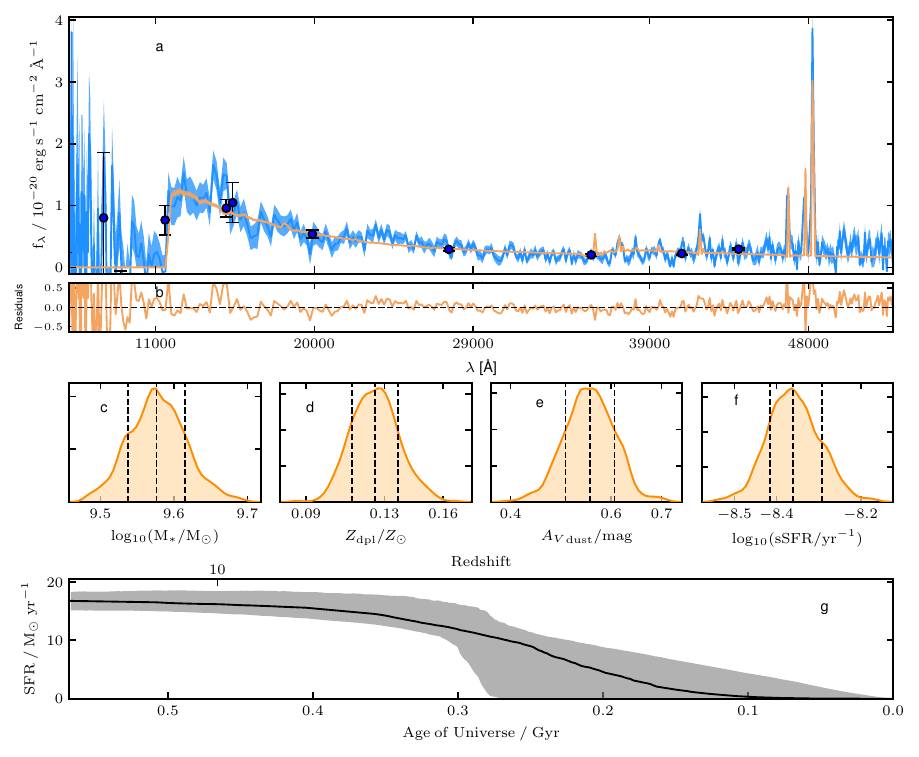}
    \caption[]{{\bf Results for the spectro-photometric SED fitting with \texttt{Bagpipes}.} Panel a: observed 1D spectrum of CANUCS-LRD-z8.6 (blue line with shaded region) and photometry (blue dots with errorbar). The \texttt{Bagpipes} best-fitting spectrum is the orange line. Panel b: residuals. Panels c,d,e,f: posterior distributions for the stellar mass, $M_*$, the metallicity, $Z$, the dust attenuation $A_{\rm v}$, the sSFR from left to right. Dashed vertical line are the 16th, 50th and 84th percentiles of the distributions. Panel g: star formation history.}
    \label{fig:sed-fitting}
\end{figure}

\begin{figure}
    \centering
    \includegraphics[width=1\linewidth]{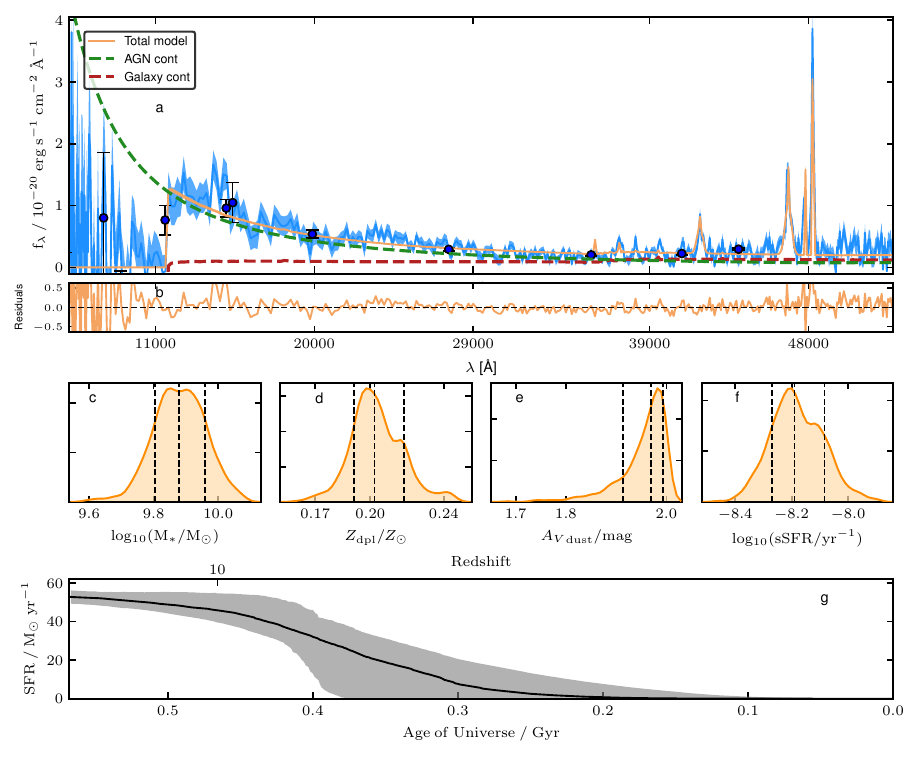}
    \caption[]{{\bf Results for the spectro-photometric SED fitting with \texttt{Bagpipes} including an AGN component.} Panel a: observed 1D spectrum of CANUCS-LRD-z8.6 (blue line with shaded region) and photometry (blue dots with errorbar). The \texttt{Bagpipes} best-fitting spectrum is the orange line, while the continuum of the AGN and of the galaxy component are shown as green and red dashed lines, respectively. Panel b: residuals. Panels c,d,e,f: posterior distributions for the stellar mass, $M_*$, the metallicity, $Z$, the dust attenuation $A_{\rm v}$, the sSFR from left to right. Dashed vertical line are the 16th, 50th and 84th percentiles of the distributions. Panel g: star formation history.}
    \label{fig:sed-fitting-AGN}
\end{figure}

\begin{figure}
    \centering
    \includegraphics[width=0.7\linewidth]{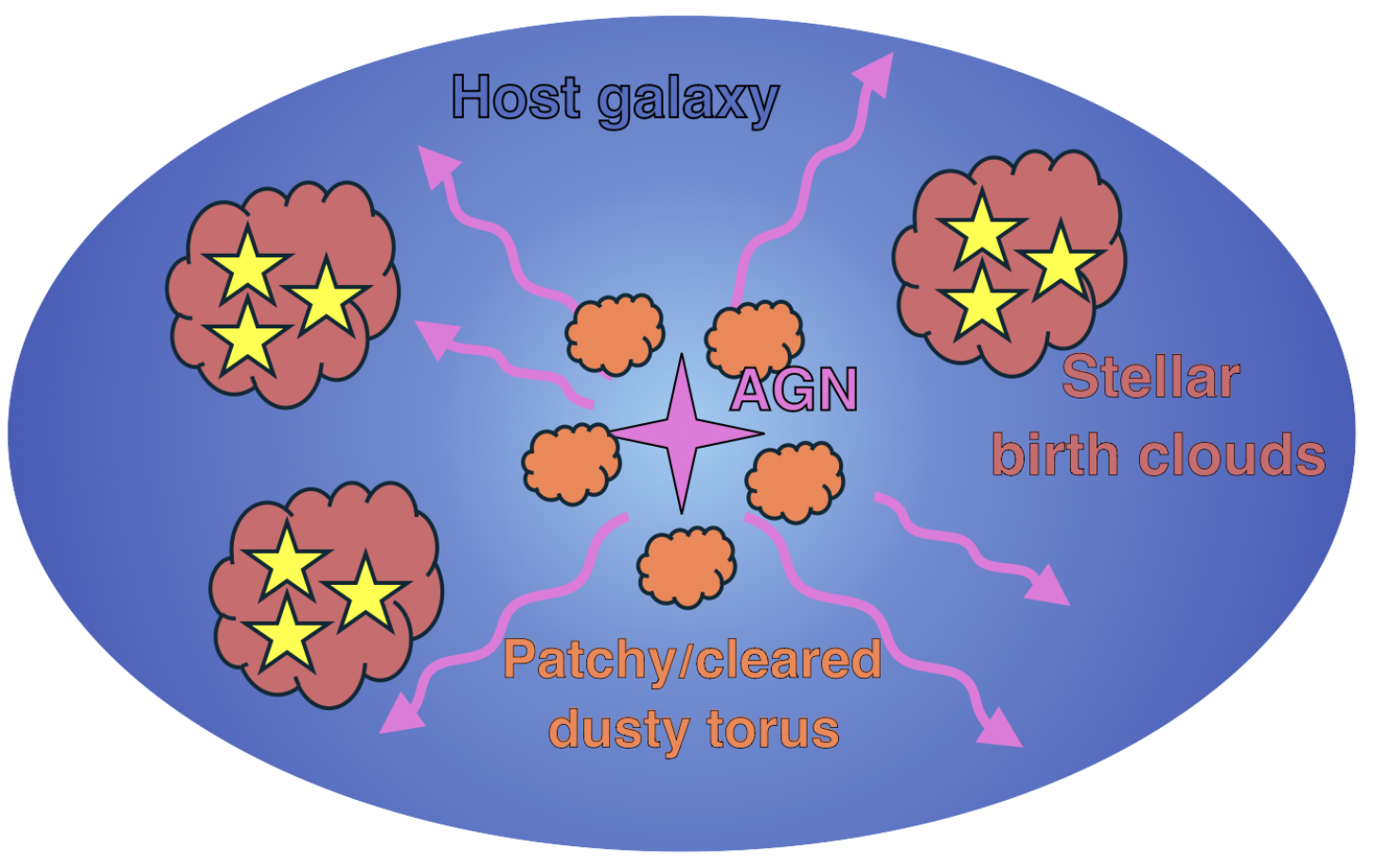}
    \caption[]{{\bf Simple visual representation for our hypothesized physical configuration of CANUCS-LRD-z8.6.} Components include a UV-bright AGN with either a patchy dusty torus or sight-line cleared by feedback. Stars are obscured by a high dust covering fraction, likely due to a combination of the current episode of star formation and the compact size.}
    \label{fig:cartoon}
\end{figure}

\begin{figure}
    \centering
    \includegraphics[width=1.\textwidth]{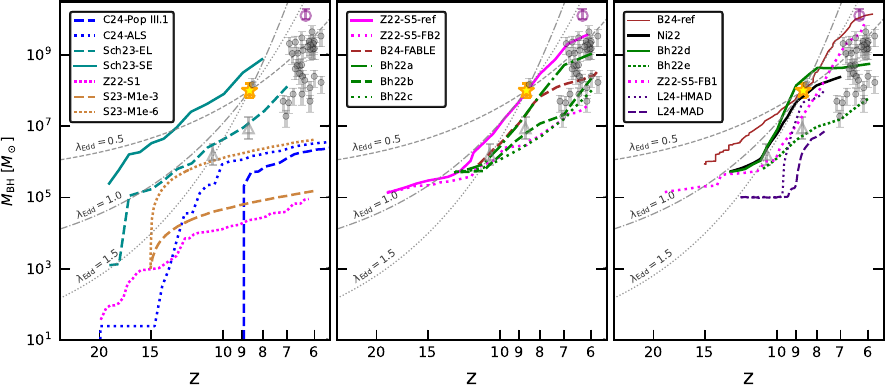} 
    \caption[]{{\bf Comparison among observations and theoretical models.} Solid (dashed or dotted) lines represent models that are (not) able to reproduce the BH mass of CANUCS-LRD-z8.6. Left: Semi-analytical models (SAM) including light and heavy seeds and hydro-dynamical simulations employing light seeds. Middle: Numerical simulations including heavy seeds, but limiting the BH accretion at the Eddington rate. Right: Numerical simulations including heavy seeds, allowing for super-Eddington growth. Details on the models included in this figure are summarised in Extended Data Table \ref{tab:models}.}
    \label{fig:models}
\end{figure}

\begin{table}[h]
        \centering
        \caption[]{Results for the line fitting of the 1D spectrum of 5112687.}
        \begin{tabular}{c|cc|cc|cc}
        \hline \hline
             &  \multicolumn{2}{c}{Fiducial} & \multicolumn{2}{c}{Test 1} & \multicolumn{2}{c}{Test 2} \\
            \hline
           Line & Flux & FWHM & Flux & FWHM & Flux & FWHM \\
            \hline
           \OIIIL[5008]  & $49.6 \pm 2.4$  & $850_{-45}^{+48}$ & $51.6 \pm 2.4$ & $870_{-45}^{+40}$ & $49.8 \pm 2.4$ & $850_{-45}^{+50}$ \\[0.2cm]
           \Hbeta$_{\rm narrow}$ & $15.6_{-2.5}^{+2.8}$ & $780_{-110}^{+120}$ & $16.2_{-2.5}^{+3.0}$ & $815_{-120}^{+110}$ & $15.3_{-2.5}^{+2.8}$ & $840_{-130}^{+95}$\\[0.2cm]
           \Hbeta$_{\rm broad}$ & $29.8 \pm 4.7$ & $4190_{-540}^{+570}$ & $26.8_{-4.7}^{+4.2}$  & $3860_{-510}^{+585}$ & $26.7 \pm 4.1$ & $3670_{-460}^{+490}$ \\[0.2cm]
           \OIIIL[4364] & $3.9 \pm 0.9$ & --$^\dagger$ & $2.4 \pm 0.9$ & --$^\dagger$  & $2.5 \pm 0.9$ & --$^\dagger$ \\[0.2cm]
           \Hgamma$_{\rm narrow}$ & $8.8 \pm 1.5$ & $950\pm 170$ & $5.0 \pm 1.4$ & $880_{-210}^{+220}$ & $5.1_{-1.1}^{+1.3}$ & $860_{-195}^{+220}$ \\[0.2cm]
           \Hgamma$_{\rm broad}$ & -- & -- & $12.5_{-2.2}^{+1.9}$ &  --$^{\dagger \dagger}$ & $9.5_{-1.9}^{+2.2}$ & --$^{\dagger \dagger}$\\[0.2cm]  
           \Hdelta & $4.0\pm 0.9$ & $1160_{-250}^{+170}$ & $4.4 \pm 0.9$ & $1220_{-220}^{+140}$ & $4.5 \pm 0.9$ & $1260_{-200}^{+115}$ \\[0.2cm]
           \NeIII & $7.1 \pm 0.9$ & $1420_{-190}^{+140}$ & $6.9 \pm 1.1$  &  $1390_{-205}^{+150}$ & $6.9 \pm 1.1$ & $1450_{-200}^{+110}$ \\[0.2cm]
           \OIIL[3727] & $<2.9$ & -- & $<2.9$ & -- & $<2.9$ & --\\[0.2cm]
           \CIV & $22.9 \pm 7.7$ & $5650_{-1500}^{+1250}$ & \multicolumn{4}{c}{as Fiducial} \\[0.2cm]
           \NIV & $21.0 \pm 7.7$ & $5670_{-1430}^{+1315}$ & \multicolumn{4}{c}{as Fiducial} \\[0.2cm]
           \hline \hline
           
        \end{tabular}
        \flushleft
        \normalsize{{\bf Notes.} Three cases are reported depending on the assumption for the broad \Hgamma component (see Sect. \Hgamma and \OIIIL[4364]). Flux is in units of $10^{-19} \rm ~erg ~s^{-1} ~cm^{-2}$. FWHM is in units of $\rm km ~s^{-1}$. $^\dagger$ FWHM$_{\OIIIL[4364]}$= FWHM$_{\OIIIL[5008]}$.  $^{\dagger \dagger}$ FWHM$_{\Hgamma_{\rm broad}}$= FWHM$_{\Hbeta_{\rm broad}}$. Upper limits are given at $3\sigma$ significance. Fluxes are not corrected for the small magnification found for the source ($\mu=1.07$, see Methods). Any magnification correction would be within the uncertainties.}
        \label{tab:line-results}
\end{table}

\begin{table}[h]
        \centering
        \caption[]{Color excess, nebular attenuation, and dust corrected ratios.}
        \begin{tabular}{c|c|c|c}
        \hline
        \hline
  & Fiducial & Test 1 & Test 2 \\
  \hline
 E(B-V)$_{\rm neb}$ & $-0.4\pm 0.5$ & $0.9\pm 0.6$ & $0.7_{-0.4}^{+0.5}$ \\[0.1cm]
           $A_{\rm V}^{\rm neb}$ & $-1.5\pm 2.0$ & $3.5\pm 2.5$ &  $2.8_{-1.8}^{+1.9}$ \\[0.1cm]
           \hline 
           Dust-corrected emission line ratios & & \\
           \hline
           $\log(\OIII/\Hbeta)$  & $0.52 \pm 0.1$ & $0.46 \pm 0.1$ & $0.48 \pm 0.1$\\[0.2cm]
           $\log(\NeIII/\OIIL[3727])$ & $>0.40$ & $>0.30$ & $>0.31$\\[0.2cm] 
           $\log(\OIIIL[4364]/\OIIIL[5008])$  & $-1.13_{-0.15}^{+0.10}$ & $-1.12_{-0.30}^{+0.10}$ & $-1.12_{-0.30}^{+0.10}$\\[0.2cm]
           $\log(\OIIIL[4364]/\Hgamma)$  & $-0.27_{-0.19}^{+0.17}$ & $-0.34\pm 0.30$ & $-0.31_{-0.30}^{+0.25}$\\[0.2cm]
           $\log(\OII/\Hbeta)$ & $<-0.92$ & $<-0.29$ & $<-0.32$ \\[0.2cm] 
           $\log(\CIV/\NIV)$ & $0.07 \pm 0.3$ & \multicolumn{2}{c}{as Fiducial} \\[0.2cm]
           \hline
           \hline

        \end{tabular}
        \flushleft
        \normalsize{{\bf Notes.} Three cases are reported depending on the assumption for the broad \Hgamma component (see Sect. \Hgamma and \OIIIL[4364]). Upper limits are given at $3\sigma$ significance.}
        \label{tab:ratios}
\end{table}

\begin{table}[h]
        \centering
        \caption[]{ AGN properties derived as described in Sect. \ref{sec:bh_prop}.}
        \begin{tabular}{c|cc}
        \hline \hline
  & Fiducial  \\
           \hline 
           $\log(M_{\rm BH, \Hbeta}/M_\odot)$ & $8.0 \pm 0.2$  \\[0.2cm]
           $\log(M_{\rm BH, 5100\AA}/M_\odot)$ & $8.2 \pm 1.3$  \\[0.2cm]
           $\log(L_{\rm Edd}/\rm erg ~s^{-1})$ & $46.2\pm 0.2$  \\[0.2cm]
           $\log(L_{\rm bol}/\rm erg ~s^{-1})$ & $45.0\pm 0.8$  \\[0.2cm]
           $\lambda_{\rm Edd}$ & 0.1  \\[0.2cm]
           \hline \hline

        \end{tabular}
        \flushleft
        \normalsize{{\bf Notes.} Quantities are corrected for magnification ($\mu=1.07$).}
        \label{tab:bh-prop}
\end{table}

\begin{sidewaystable}[h]
    \centering
    \caption[]{Model parameters and results for the spectro-photometric SED fitting with \texttt{Bagpipes}.}
    \begin{tabular}{c|cccc}
    \hline \hline
    Parameter & Range & Prior & Result w/o AGN & Result with AGN \\
    \hline
      dblplaw $\tau$ & (0., 15.) & flat & $0.30_{-0.06}^{+0.14}$ & $0.41_{-0.06}^{+0.08}$\\[0.1cm]
      dblplaw $\alpha$ & (0.01, 1000.) & log10 & $0.07_{-0.05}^{+0.26}$ & $0.08_{-0.06}^{+0.37}$ \\[0.1cm]
      dblplaw $\beta$ & (0.01, 1000.) & log10 & $4_{-3}^{+165}$ & $10.8_{-7.9}^{+247}$\\[0.1cm]
      $\log(M_{\rm *}/M_\odot)$  & (6., 13.) & flat & $9.58_{-0.04}^{+0.03}$ & $9.88\pm 0.08$\\[0.1cm]
      dblplaw $Z/Z_\odot$ & (0.01, 0.25) & flat & $0.13\pm 0.01$ & $0.20\pm 0.01$ \\[0.1cm]
      A$_{\rm V}^{\rm star}$ & (0., 2.0) & flat & $0.56\pm 0.05$ & $1.97_{-0.05}^{+0.02}$ \\[0.1cm]
      $\log(\rm U)$ & -1.5 & -- & -- & --\\
      agn $\alpha_\lambda$ & (-2., 2.) & gaussian: $\mu=-1.5, \sigma=0.5$ & -- & $-1.99\pm 0.01$\\
      agn $\beta_\lambda$ & (-2., 2.) & gaussian: $\mu=0.5, \sigma=0.5$ & -- & $0.75\pm 0.50$\\
      agn ${\rm H}\alpha_{\rm norm}$ & (0., $25.\times 10^{-18}$) & flat & -- & $(1.08\pm 0.08) \times 10^{-17}$\\
      agn $F_{5100\AA}$ & (0., $10^{-19}$) & flat & -- & $(7.1\pm 0.3) \times 10^{-21}$\\
      agn $\sigma$ & (1000., 5000.) & log10 & -- & $1139_{-82}^{+133}$\\[0.1cm]

      \hline \hline
    \end{tabular}
    \flushleft
    \normalsize{{\bf Notes.} Quantities are not corrected for magnification.}
    \label{tab:res-bagpipes}
\end{sidewaystable}

\begin{sidewaystable}[h]
        \centering
        \caption[]{Summary of the main assumptions adopted in the SAM and numerical simulations considered.} 
        \begin{tabular}{c|c|c|c|c}
        \hline
        \hline
  Name & Accretion and feedback properties & $M_{\rm seed}$ [$M_\odot$] & $\lambda_{\rm Edd}$ & ref.  \\
  \hline
           C24-Pop III.1 & GAEA semi-analytic model \cite{fontanot2020} & $10^5$  & 10  & \cite{Cammelli24} \\[0.2cm]
           C24-ALS  & GAEA semi-analytic model \cite{fontanot2020} & dependent on the halo mass & 10 & \cite{Cammelli24} \\[0.2cm]
           Sch23-EL & $\alpha$ free, $\epsilon=0.1$ & Light seeds: $40-300$ + Heavy seeds: $10^5$ & $1$ & \cite{Schneider2023} \\[0.2cm]
           Sch23-SE & $\alpha$ free, slim disc & Light seeds: $40-300$ + Heavy seeds: $10^5$ & SE & \cite{Schneider2023} \\[0.2cm]
           S23-M1e-6 & $\alpha=1$, slim disc, low feedback coupling & $10^{3}$ & SE  & \cite{Sassano2023} \\ [0.2cm]
           S23-M1e-3 & $\alpha=1$, slim disc, high feedback coupling & $10^{3}$ & SE  & \cite{Sassano2023}\\[0.2cm]
           Z22-S1 & $\alpha=100$, slim disc & $10 \ h^{-1}$ & $10^{4}$  & \cite{zhu2022} \\[0.2cm] 
           Z22-S5-ref & $\alpha=100$, $\epsilon=0.1$ & $10^{5} \ h^{-1}$ &  $1$ & \cite{zhu2022}\\[0.1cm]
           Z22-S5-FB1 & $\alpha=100$, slim disc & $10^{5} \ h^{-1}$ &  $10^{4}$  & \cite{zhu2022} \\[0.2cm]
           Z22-S5-FB2 & $\alpha=100$, $\epsilon=0.2$ & $10^{5} \ h^{-1}$ &  $1$  & \cite{zhu2022}\\[0.2cm]
           B23-FABLE & $\alpha=100$, $\epsilon=0.1$ & $10^{5} \ h^{-1}$ &  $1$ & \cite{bennett2024} \\[0.2cm]
           B23-ref & $\alpha=100$, $\epsilon=0.1$, reduced feedback coupling  & $10^{5} \ h^{-1}$ & $2$  & \cite{bennett2024} \\[0.2cm]
           Bh22a & $\alpha=100$, $\epsilon=0.1$ & $5 \times 10^{5} \ h^{-1}$ &  $1$ & \cite{Bhowmick2022}\\[0.2cm]
           Bh22b &$\alpha=1$, $\epsilon=0.1$ & $5 \times 10^{5} \ h^{-1}$ &  $1$  & \cite{Bhowmick2022}\\[0.2cm]
           Bh22c & $\alpha=100$, $\epsilon=0.2$ & $5 \times 10^{5} \ h^{-1}$ &  $1$  & \cite{Bhowmick2022}\\[0.2cm]
           Bh22d & $\alpha=100$, $\epsilon=0.1$ & $5 \times 10^{5} \ h^{-1}$ &  $2$  & \cite{Bhowmick2022}\\[0.2cm]
           Bh22e & $\alpha=1$, $\epsilon=0.2$ & $5 \times 10^{5} \ h^{-1}$ &  $2$  & \cite{Bhowmick2022}\\[0.2cm]
           Ni22 & $\alpha=100$, $\epsilon=0.1$ & $5 \times 10^{5} \ h^{-1}$ &  $2$ & \cite{Ni2022} \\[0.2cm]
           L24-HMAD & $\alpha=1$, several feedback regimes  & $10^{5}$ & SE & \cite{Lupi2024} \\[0.2cm]
           L24-MAD & $\alpha=1$, several feedback regimes & $10^{5}$ &  SE & \cite{Lupi2024} \\[0.2cm]
           \hline
           \hline
        \end{tabular}
        \flushleft
        \normalsize{{\bf Notes.} This description is not complete by any means and we refer to the original works for a complete overview of the numerical setup. \textit{First column:} Name of the model, as reported in Extended Data Figure \ref{fig:models}. \textit{Second column:} Relevant parameters in the accretion rate modeling and AGN feedback implementation. In particular, we emphasize the boost factor $\alpha$ of the accretion rate and the radiative efficiency $\epsilon$. In models employing the slim disc solution the radiative efficiency depends on the accretion rate as described in \cite{Madau2001}. \textit{Third column:} maximum Eddington ratio $\lambda_{\rm Edd}$ considered. $\lambda_{\rm Edd}=1$ implies Eddington-limited growth, SE labels runs where no limit is considered. \textit{Fourth column:} Model reference.}
        \label{tab:models}
\end{sidewaystable}
\clearpage
\setcounter{figure}{0}   
\setcounter{table}{0}

\end{document}